\documentclass[manuscript, noblind]{geophysics}

\usepackage{amsmath}
\usepackage{subcaption}
\usepackage{url}
\begin{document}

\title{Upside down Rayleigh-Marchenko: a practical, yet exact redatuming scheme for seabed seismic acquisitions}

\renewcommand{\thefootnote}{\fnsymbol{footnote}} 

\address{
Division of Earth Science and Engineering (ErSE), 
King Abdullah University of Science and Technology (KAUST),
23955-6900, Thuwal, Saudi Arabia,
{ning.wang.2, matteo.ravasi}@kaust.edu.sa}
\author{Ning Wang and Matteo Ravasi}

\footer{Example}
\lefthead{Wang \& Ravasi}
\righthead{Upside down Rayleigh-Marchenko redatuming}


\begin{abstract}
Ocean-bottom seismic plays a crucial role in resource exploration and monitoring. However, despite its undoubted potential, the use of coarse receiver geometries poses challenges to accurate wavefield redatuming. This in turn, affects the quality of subsequent imaging and reservoir charactherization products. We propose a reciprocal version of the Rayleigh-Marchenko method, called upside down Rayleigh-Marchenko, where all spatial integrals are performed over the (usually much better-sampled) source carpet; this results in a theoretically exact redatuming scheme, which can handle irregular and sparse receiver geometries. The proposed method requires availability of multi-component receivers and either dual-sensor sources or a pre-processing step of model-based source deghosting, and utilizes only the down-going component of the receiver-side wavefield; as such, it can be interpreted as a full-wavefield extension of the mirror imaging method commonly used in seabed settings. Two synthetic examples are used to showcase the effectiveness of the proposed method, starting from the ideal scenario of finely and regularly sampled sources and receivers, and later considering different levels of decimation for the receiver array. Migrated images as well as common-angle gathers reveal that our method can be used to produce structural and amplitude-friendly imaging outputs with minimal data pre-processing.
\end{abstract}

\section{Introduction}
Ocean-bottom seismic (OBS) represents a cost-effective, high-quality acquisition solution to image and monitor the subsurface for applications ranging from hydrocarbon exploration and management, $\mathrm{CO}_2$ sequestration and monitoring, geothermal production, and de-risking of offshore wind farms. However, due to the expensive nature of seismic nodes and the associated costs for the deployment and retrieval of such an equipment along the seabed, OBS surveys are routinely acquired with coarse receiver coverage (in the crossline direction for ocean-bottom cables, and in both directions for ocean-bottom nodes). This poses challenges to achieving accurate wavefield redatuming and reliable imaging.

To deal with such shortcomings, \cite{Grion2007} proposed mirror imaging as a way to increase the illumination of OBS data, especially for shallow targets, by using down-going ghost reflections as a complement to the more commonly imaged up-going primary reflections. Since then, the underlying idea interchanging the role of sources and receivers in imaging has been adopted in other wave-equation-based imaging algorithms to deal with the presence of sparse receivers, leveraging the usually available denser grid of sources for wavefield extrapolation. For example, \cite{Lu2015} developed an imaging algorithm able to handle both primaries and free-surface multiples that only utilizes the down-going component of the recorded data. Despite being practically attractive, such an approach relies on single-scattering assumptions and therefore is prone to the generation of cross-talk imaging artifacts~\citep{Lu2016}.

Marchenko redatuming \citep{Broggini2012, Wapenaar2013, Slob2014, Wapenaar2014, Broggini2014, vanderNeut2015, Wapenaar2017} is a data-driven method that aims to retrieve up- and down-going Green’s functions between a virtual source (also called focal point) and receivers positioned at the surface of the Earth. This is achieved by solving the so-called Marchenko equations. The down-going focusing function can be interpreted as a wavefield, which once injected at the surface of the Earth produces a focus at the defined focal point; the up-going focusing function contains instead the waves reflected off interfaces from the down-going focusing function and recorded at the surface \citep{Wapenaar2014}. As the Marchenko method can retreive full-wavefield Green’s functions, which include accurate internal multiple reflections, resulting imaging products are deprived of artifacts related to such multiples. Several successful field data applications of the Marchenko redatuming method have been presented to date \citep{Ravasi2015,Ravasi2016,Jia2017,Staring2017,Staring2020, Staring2021}, proving its ability to deal with multiples in both simple and complex geological scenarios. Nevertheless, tailored pre-processing sequences have been carried out in these studies to accommodate for the strict data requirements of the Marchenko scheme. To overcome this limitation, \cite{Ravasi2017} introduced the Rayleigh-Marchenko (RM) method, combining the one-way version of the Rayleigh integral representation \citep{Wapenaar1989} with the Marchenko equations. By doing so, the RM method accomodates for the presence of an inhomogeneous boundary condition above the acquisition level (i.e., correctly handling free-surface multiples), relaxes the requirement for co-location of sources and receivers, and eliminates the need for an accurate estimation and deconvolution of the source signature from the reflection data. On the other hand, the RM method requires recorded data to be decomposed in their up- and down-going components on the receiver-side, a step that is commonly performed when processing multi-component OBS data. Despite its great potential, the RM method still requires a dense and finely sampled receiver array. This clearly represents a limitation when considering novel seabed seismic acquisition geometries with sparse receiver arrays.

In this paper, we propose the Upside Down Rayleigh-Marchenko (UD-RM) method, the first practical, yet exact redatuming scheme for ocean-bottom data that can theoretically accommodate for sparse receiver grids. This is achieved by invoking reciprocity to interchange the role of sources and receivers. By doing so, all spatial integrals in the new set of Marchenko equations are applied along the source side. Moreover, since receivers are mirrored with respect to the sea surface as in mirror imaging, the UD-RM method is expected to provide subsurface wavefields and images with expanded illumination. In terms of data pre-processing, UD-RM requires the recorded data to be separated into up- and down-going components on both the source and receiver sides. More specifically, the receiver-side down-going wavefield, separated into its up- and down-going components on the source-side and deprived of the direct wave, represents our input data. As such, the proposed methodology requires availability of multi-component receivers and dual-sensor sources \citep{Vasconcelos2013}. Alternatively, a step of source-side deghosting can be used to accommodate for the presence of single-sensor sources; this reduces the cost of acquiring data and increases the applicability of the proposed method. In our numerical example, a deghosting method similar to that of \cite{Grion2016} is applied; however, any other deghosting method can be equivalently used as later explained in the Discussion section. Interestingly, we can interpret the approach of \cite{Lu2015} as the application of our operator to single-scattering propagators (i.e., initial focusing functions) in the subsurface.

In the following, we briefly recall the theory of Marchenko and RM redatuming and present the theory of the proposed method. This is followed by a validation of its numerical effectiveness on two synthetic examples of increasing complexity: first, we consider a constant-velocity, variable-density model consisting of a simple layered geology with a high density syncline structure; second, a 2D section of the EAGE/SEG Overthrust model \citep{Aminzadeh1997} with an additional water layer is considered. In both cases, redatuming and imaging is initially performed with a dense receiver grid; subsequent tests are presented to assess the ability of the proposed methodology to handle geometries with increased spacing between receivers. In the latter case, two different optimizers are considered when solving the UD-RM equations: the same least-squares solver used to estimate the focusing functions in the case of dense receiver grids, and, similar to \cite{Haindl2021}, a sparsity promoting solver (more precisely, the Fast Iterative Soft Thresholding algorithm - FISTA, \cite{Beck2009}) with a sliding linear Radon sparsifying transform. Despite its increased computational cost, the latter approach is shown to produce images and common-image gathers of comparable quality to those obtained from a dense receiver geometry (with up to 60\% of missing receivers).

\section*{Theory}
\subsection{Marchenko equations}
\begin{figure*}
  \centering
  \includegraphics[width=0.92\textwidth]{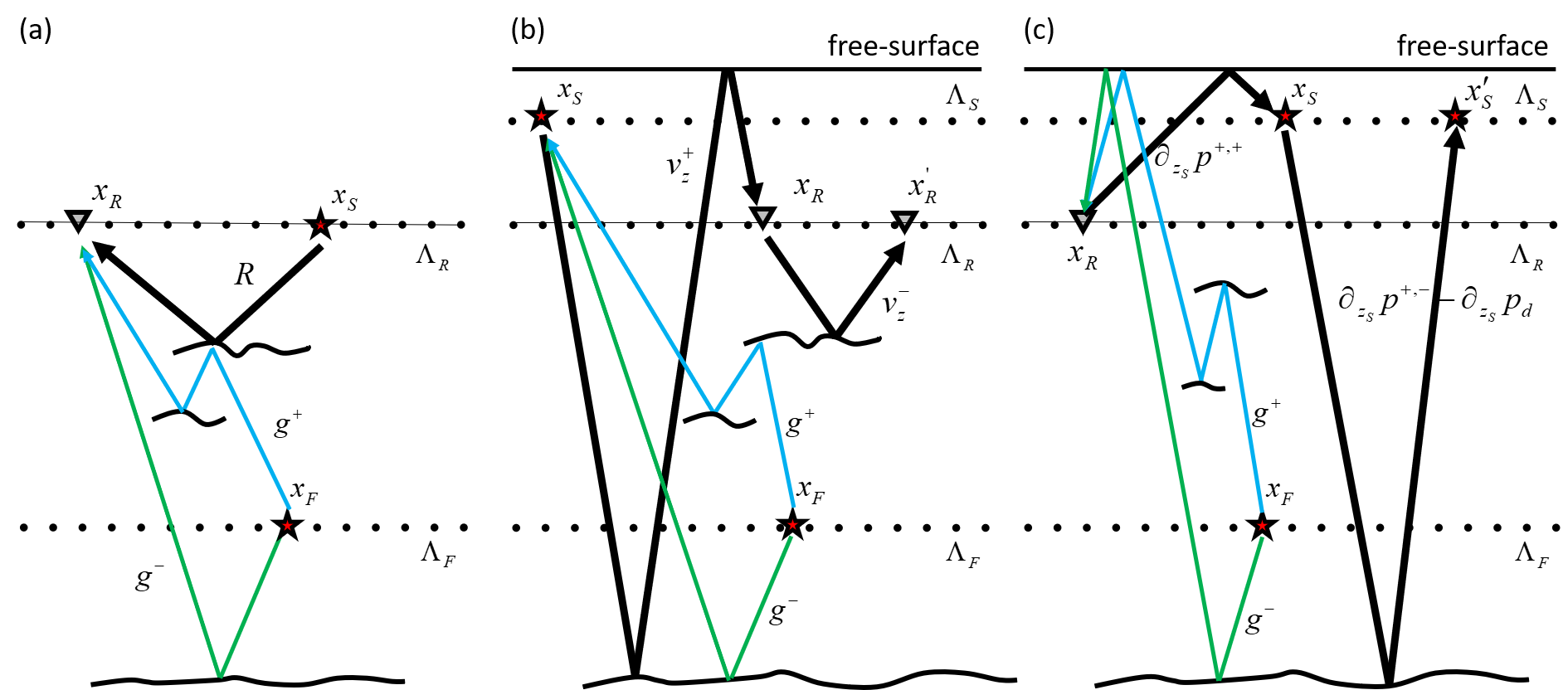}
  \caption{Geometries of (a) the original Marchenko method, (b) the RM method, (c) the UD-RM method.}
  \label{fig:geometries}
\end{figure*}

Let us consider a heterogeneous, lossless medium below a transparent boundary (i.e., acquisition level $\Lambda_{R}$) and a second boundary at a depth level $\Lambda_{F}$ (i.e., focal point or virtual source level) as shown in Figure \ref{fig:geometries}a. Given the surface reflection response ${R}(\textbf{x}_{R},\textbf{x}_{R}')$ from sources at $\textbf{x}_{R}'$ placed just above the acquisition level $\Lambda_{R}$ to receivers at $\textbf{x}_{R}$ and the up- and down-going focusing functions ($f^{-}(\textbf{x}_{R},\textbf{x}_{F})$ and $f^{+}(\textbf{x}_{R},\textbf{x}_{F})$), the up- and down-going Green's functions from the focal point to receivers ($g^{-}(\textbf{x}_{F},\textbf{x}_{R})$ and $g^{+}(\textbf{x}_{F},\textbf{x}_{R})$) can be expressed as~\citep{Wapenaar2014, vanderNeut2015}
\begin{equation}
\label{eq:Marchenko_gup}
g^{-}(\textbf{x}_{F},\textbf{x}_{R}) = \int _{\Lambda _{R}}R(\textbf{x}_{R},\textbf{x}_{R}')f^{+}(\textbf{x}_{R}',\textbf{x}_{F})d\textbf{x}_{R}' - f^{-}(\textbf{x}_{R},\textbf{x}_{F})
,
\end{equation}
\begin{equation}
\label{eq:Marchenko_gdown}
-g^{+*}(\textbf{x}_{F},\textbf{x}_{R}) = \int _{\Lambda _{R}}R^{*}(\textbf{x}_{R},\textbf{x}_{R}')f^{-}(\textbf{x}_{R}',\textbf{x}_{F})d\textbf{x}_{R}' - f^{+}(\textbf{x}_{R},\textbf{x}_{F})
,
\end{equation}
where superscripts - and + are used to define the up- and down-going components, while * represents complex conjugation in the frequency domain (or time reversal in time domain); the reflection response ${R}$ can be expressed as~\citep{Wapenaar1989}
\begin{equation}
\label{eq:R}
{R}(\textbf{x}_{R},\textbf{x}_{R}') = \frac{2}{j\omega \rho(\textbf{x}_{R}')}\partial _{z'}g_{0}(\textbf{x}_{R},\textbf{x}_{R}')
,
\end{equation}
where $j$, $\omega$, and $\rho$ are the imaginary unit, angular frequency, and density at $\textbf{x}_{R}'$, respectively. Here, $g_{0}$ represents the total pressure wavefield from a monopole source without free-surface effects and deprived the of direct wave, while $\partial _{z'}$ is the derivative along the vertical axis at $\textbf{x}_{R}'$.

\subsection{Rayleigh-Marchenko equations}
The RM method~\citep{Ravasi2017} includes heterogeneities and/or a different boundary condition above the acquisition level $\Lambda_{R}$ (e.g., free-surface) by coupling the Marchenko equations with the following one-way version of the Rayleigh integral representation~\citep{Wapenaar1989}
\begin{equation}
\label{eq:Rayleigh_integral}
-v_{z}^{-}(\textbf{x}_{R}',\textbf{x}_{S}) = \int _{\Lambda _{R}}v_{z}^{+}(\textbf{x}_{R},\textbf{x}_{S})R(\textbf{x}_{R},\textbf{x}_{R}')d\textbf{x}_{R}
,
\end{equation}
where $v_{z}^{-}$ and $v_{z}^{+}$ represent the up- and down-going vertical particle velocities, respectively, recorded from sources at $\textbf{x}_{S}$ along $\Lambda_{S}$ and separated at $\textbf{x}_{R}$ along $\Lambda_{R}$. The RM equations can be obtained by pre-multiplying $v_{z}^{+}(\textbf{x}_{R},\textbf{x}_{S})$ and $v_{z}^{+*}(\textbf{x}_{R},\textbf{x}_{S})$ to the original Marchenko equations \ref{eq:Marchenko_gup} and \ref{eq:Marchenko_gdown}, respectively, and applying integration along the receiver array
\begin{equation}
\begin{aligned}
\label{eq:RM_pup}
p^{-}(\textbf{x}_{F},\textbf{x}_{S}) = &-\int _{\Lambda _{R}}v_{z}^{-}(\textbf{x}_{R}',\textbf{x}_{S})f^{+}(\textbf{x}_{R}',\textbf{x}_{F})d\textbf{x}_{R}'\\ &-\int _{\Lambda _{R}} v_{z}^{+}(\textbf{x}_{R},\textbf{x}_{S})f^{-}(\textbf{x}_{R},\textbf{x}_{F})d\textbf{x}_{R}
\end{aligned}
 ,
\end{equation}
\begin{equation}
\begin{aligned}
\label{eq:RM_pdown}
p^{+*}(\textbf{x}_{F},\textbf{x}_{S}) = &\int _{\Lambda _{R}}v_{z}^{-*}(\textbf{x}_{R}',\textbf{x}_{S})f^{-}(\textbf{x}_{R}',\textbf{x}_{F})d\textbf{x}_{R}'\\ &+\int _{\Lambda _{R}} v_{z}^{+*}(\textbf{x}_{R},\textbf{x}_{S})f^{+}(\textbf{x}_{R},\textbf{x}_{F})d\textbf{x}_{R}
\end{aligned}
,
\end{equation}
where $p^{-/+}$ is the band-limited up-/down-going pressure wavefield that includes surface-related multiples
\begin{equation}
\label{eq:pressure_wavefield}
p^{-/+}(\textbf{x}_{F},\textbf{x}_{S}) = \int _{\Lambda _{R}}v_{z}^{+}(\textbf{x}_{R},\textbf{x}_{S})g^{-/+}(\textbf{x}_{F},\textbf{x}_{R})d\textbf{x}_{R}
.
\end{equation}

Equations \ref{eq:RM_pup} and \ref{eq:RM_pdown} reveal that the RM method can include the effects of the free-surface and allow sources and receivers to be placed at different depth levels (Figure \ref{fig:geometries}b), a condition required to be able to apply such a method to OBS acquisition geometries. Nevertheless, all spatial integrals are still applied along the $\Lambda_{R}$ level, meaning that the RM method requires a dense and finely sampled receiver array.

\subsection{Upside-down Rayleigh-Marchenko equations}
In practical OBS acquisition systems, the requirement of densely sampled receiver arrays is rarely met. Therefore, we propose here to modify the RM method by leveraging reciprocity to interchange the role of sources and receivers; this ensures that all spatial integrals are now carried out over the source array. To begin with, we rewrite the Marchenko equations from sources at $\textbf{x}_{S}$ and receivers at $\textbf{x}_{S}'$ at the same depth level
\begin{equation}
\label{eq:Marchenko_gup_swap}
g^{-}(\textbf{x}_{F},\textbf{x}_{S}) = \int _{\Lambda _{S}}R(\textbf{x}_{S},\textbf{x}'_{S})f^{+}(\textbf{x}_{S}',\textbf{x}_{F})d\textbf{x}_{S}' - f^{-}(\textbf{x}_{S},\textbf{x}_{F})
,
\end{equation}
\begin{equation}
\label{eq:Marchenko_gdown_swap}
-g^{+*}(\textbf{x}_{F},\textbf{x}_{S}) = \int _{\Lambda _{S}}R^{*}(\textbf{x}_{S},\textbf{x}_{S}')f^{-}(\textbf{x}_{S}',\textbf{x}_{F})d\textbf{x}_{S}' - f^{+}(\textbf{x}_{S},\textbf{x}_{F})
,
\end{equation}
where the reflection response ${R}$ is defined as in equation \ref{eq:R} with different sources and receivers.

Let us now consider a receiver at $\textbf{x}_{R}$ and a virtual receiver at $\textbf{x}_{S}'$ inside an enclosing boundary with upper boundary at $\Lambda _{S}$ and vanishing lower boundary because of the Sommerfeld radiation condition. If we identify a first state with the same properties of the physical medium and free-surface (represented by $g$) and a second, ideal medium without free-surface (represented by $g_{0}$), the following representation theorem holds true ~\citep{Almagro2014}
\begin{equation}
\begin{aligned}
\label{eq:g_g0}
g^{.,-}(\textbf{x}_{R},\textbf{x}_{S}') &- g_{0}(\textbf{x}_{R},\textbf{x}_{S}') = 
\\&-\int _{\Lambda _{S}}\dfrac{2}{j\omega \rho(\textbf{x}_{S})}\partial _{z}g^{.,+}(\textbf{x}_{R},\textbf{x}_{S})g_{0}(\textbf{x}_{S},\textbf{x}_{S}')d\textbf{x}_{S}
\end{aligned}
,
\end{equation}
where the first element of the superscript $^{.,.}$ represents the wavefield component at the receiver side and the second identifies the component at the source side, $\partial_{z}$ corresponds to the vertical derivative at $\textbf{x}_{S}$. In this work, we follow the convention that wavefields reaching a source (or a receiver) from above have a plus (+) superscript, whilst those reaching them from below have a minus (-) superscript. After some manipulation (see Appendix A for the full derivation), we obtain
\begin{equation}
\begin{aligned}
\label{eq:UD_g}
g^{+,-}(\textbf{x}_{R},\textbf{x}_{S}') &- g_{d}(\textbf{x}_{R},\textbf{x}_{S}') = 
\\&-\int _{\Lambda _{S}}\dfrac{2}{j\omega \rho(\textbf{x}_{S})}\partial _{z}g^{+,+}(\textbf{x}_{R},\textbf{x}_{S})g_{0}(\textbf{x}_{S},\textbf{x}_{S}')d\textbf{x}_{S}
\end{aligned}
,
\end{equation}
where $g_{d}(\textbf{x}_{R},\textbf{x}_{S}')$ represents the direct wave from a source at $\textbf{x}_{S}'$ to a receiver at $\textbf{x}_{R}$. In order to be able to combine equation \ref{eq:UD_g} with the Marchenko equations \ref{eq:Marchenko_gup_swap} and \ref{eq:Marchenko_gdown_swap}, we multiply each side by $\dfrac{s(\omega)}{j\omega\rho(\textbf{x}_{S}')}\partial _{z'}$:
\begin{equation}
\begin{aligned}
\label{eq:UD_g_multiply_s}
&\dfrac{s(\omega)}{j\omega\rho(\textbf{x}_{S}')}\partial _{z'}(g^{+,-}(\textbf{x}_{R},\textbf{x}_{S}') - g_{d}(\textbf{x}_{R},\textbf{x}_{S}')) = 
\\&\int _{\Lambda _{S}}-\dfrac{2}{j\omega \rho(\textbf{x}_{S})}\partial _{z}g^{+,+}(\textbf{x}_{R},\textbf{x}_{S}) \left(\dfrac{s(\omega)}{j\omega\rho(\textbf{x}_{S}')}\partial _{z'}g_{0}(\textbf{x}_{S},\textbf{x}_{S}')\right)d\textbf{x}_{S}
\end{aligned}
,
\end{equation}
where $s$ represents the source wavelet. From equation \ref{eq:R}, we can recognize that the last term within parenthesis on the right hand side of this equation corresponds to the reflection response along the source level $\Lambda_{S}$. As such, we can write an equation linking $R$ to the source-side separated, down-going wavefields at the receiver side and the direct wave:
\begin{equation}
\begin{aligned}
\label{eq:UD_p}
&\partial _{z}p^{+,-}(\textbf{x}_{R},\textbf{x}_{S}') - \partial _{z}p_{d}(\textbf{x}_{R},\textbf{x}_{S}')= 
\\&-\int _{\Lambda _{S}}\partial _{z}p^{+,+}(\textbf{x}_{R},\textbf{x}_{S}) R(\textbf{x}_{S},\textbf{x}_{S}')d\textbf{x}_{S}
\end{aligned}
,
\end{equation}
where $p$ represents the pressure wavefield. Equation \ref{eq:UD_p} effectively represents a source-side equivalent of the well-known free-surface demultiple method by multi-dimensional deconvolution \citep{Hampson2020, Boieroetal2023, Haacke2023}. Considering the geometry in Figure \ref{fig:geometries}c, which is identical to that in Figure \ref{fig:geometries}b except for the fact that sources and receivers are interchanged, the physical meaning of equation \ref{eq:UD_p} can be explained as follows: the down-going multiples are used here as an aerial source wavefield that, once convolved with the reflection response between sources, give rise to the total up-going wavefield from receivers to sources minus the direct arrival.

Akin to the RM method, after multiplying each side of equation \ref{eq:Marchenko_gup_swap} by the mirrored down-going multiples $\int _{\Lambda _{S}}\partial _{z}p^{+,+}(\textbf{x}_{R},\textbf{x}_{S})$ and each side of equation \ref{eq:Marchenko_gdown_swap} by its complex conjugated $\int _{\Lambda _{S}}\partial _{z}p^{+,+*}(\textbf{x}_{R},\textbf{x}_{S})$, and utilizing equation \ref{eq:UD_p}, we obtain a set of equations whose spatial integrals are now performed only over the source coordinates:
\begin{equation}
\begin{aligned}
\label{eq:UDRM_up}
g^{+,-}(\textbf{x}_{R},\textbf{x}_{F}) = &-\int _{\Lambda _{S}}\partial _{z}\widetilde{p}^{+,-}(\textbf{x}_{R},\textbf{x}_{S}') f^{+}(\textbf{x}_{S}',\textbf{x}_{F})d\textbf{x}_{S}' 
\\&- \int _{\Lambda _{S}}\partial _{z}p^{+,+}(\textbf{x}_{R},\textbf{x}_{S}) f^{-}(\textbf{x}_{S},\textbf{x}_{F})d\textbf{x}_{S}
\end{aligned}
,
\end{equation}
\begin{equation}
\begin{aligned}
\label{eq:UDRM_down}
-g^{+,+*}(\textbf{x}_{R},\textbf{x}_{F}) = &-\int _{\Lambda _{S}}\partial _{z}\widetilde{p}^{+,-*}(\textbf{x}_{R},\textbf{x}_{S}') f^{-}(\textbf{x}_{S}',\textbf{x}_{F})d\textbf{x}_{S}' 
\\&- \int _{\Lambda _{S}}\partial _{z}p^{+,+*}(\textbf{x}_{R},\textbf{x}_{S}) f^{+}(\textbf{x}_{S},\textbf{x}_{F})d\textbf{x}_{S}
\end{aligned}
,
\end{equation}
where the terms on the left-hand side are equivalent to the up- and down-going subsurface wavefields including free-surface effects from the focal point to the available receivers (only down-going component at receivers):
\begin{equation}
\label{eq:g+-}
g^{+,-}(\textbf{x}_{R},\textbf{x}_{F}) = \int _{\Lambda _{S}}\partial _{z}p^{+,+}(\textbf{x}_{R},\textbf{x}_{S}) g^{-}(\textbf{x}_{S},\textbf{x}_{F})d\textbf{x}_{S}
,
\end{equation}
\begin{equation}
\label{eq:g++*}
g^{+,+*}(\textbf{x}_{R},\textbf{x}_{F}) = \int _{\Lambda _{S}}\partial _{z}p^{+,+*}(\textbf{x}_{R},\textbf{x}_{S}) g^{+*}(\textbf{x}_{S},\textbf{x}_{F})d\textbf{x}_{S}
,
\end{equation}
while $\partial _{z}\widetilde{p}^{+,-}(\textbf{x}_{R},\textbf{x}_{S}')$ represents the source-side up-going wavefield deprived of the direct wave:
\begin{equation}
\label{eq:widetilde_p}
\partial _{z}\widetilde{p}^{+,-}(\textbf{x}_{R},\textbf{x}_{S}') = \partial_{z}p^{+,-}(\textbf{x}_{R},\textbf{x}_{S}') - \partial_{z}p_d(\textbf{x}_{R},\textbf{x}_{S}')
.
\end{equation}
Equations \ref{eq:UDRM_up} and \ref{eq:UDRM_down} can be rewritten in a matrix form:
\begin{equation}
\label{eq:UDRM_matrix}
\begin{bmatrix}
    \mathbf{-g^{+,-}} \\
    \mathbf{g^{+,+*}}
\end{bmatrix}
=
\begin{bmatrix}
    \partial_{z}\mathbf{P}^{+,+} & \partial_{z}\widetilde{\mathbf{P}}^{+,-} \\
    \partial_{z}\widetilde{\mathbf{P}}^{+,-*} & \partial_{z}\mathbf{P}^{+,+*} 
\end{bmatrix}
\begin{bmatrix}
    \mathbf{f^{-}} \\
    \mathbf{f^{+}}
\end{bmatrix}
.
\end{equation}

In order to be able to solve equation \ref{eq:UDRM_matrix} for the focusing functions, we introduce a windowing operator $\Theta_{-x_R}$, which removes all the events after the traveltime $t_{d}(-\textbf{x}_{R},\textbf{x}_{F})$ of the first arriving wave from the focal point $\textbf{x}_{F}$ to the mirror receivers $-\textbf{x}_{R}$ (i.e., a wave that propagates from the focal point to the free-surface and bounces off towards the receiver) as well as those before the negative traveltime $-t_{d}(-\textbf{x}_{R},\textbf{x}_{F})$. Since the Green’s functions contain events only after the traveltime of the first arriving wave and before the negative traveltime, the up- and down-going subsurface wavefields are zeroed by the windowing operator (i.e., $\Theta_{-x_R}\mathbf{g^{+,-}}=\mathbf{0}$ and $\Theta_{-x_R}\mathbf{g^{+,+}}=\mathbf{0}$). The resulting system of equations becomes:
\begin{equation}
\label{eq:UDRM_matrix_window}
\begin{bmatrix}
    -\Theta_{-x_R}\partial_{z}\widetilde{\mathbf{P}}^{+,-}\mathbf{f}_\mathbf{d}^+ \\
    -\Theta_{-x_R}\partial_{z}\mathbf{P^{+,+*}}\mathbf{f}_\mathbf{d}^+
\end{bmatrix}
=
\begin{bmatrix}
    \Theta_{-x_R}\partial_{z}\mathbf{P}^{+,+} & \Theta_{-x_R}\partial_{z}\widetilde{\mathbf{P}}^{+,-} \\
    \Theta_{-x_R}\partial_{z}\widetilde{\mathbf{P}}^{+,-*} & \Theta_{-x_R}\partial_{z}\mathbf{P}^{+,+*} 
\end{bmatrix}
\begin{bmatrix}
    \mathbf{f^{-}} \\
    \mathbf{f^{+}_m}
\end{bmatrix}
,
\end{equation}
where $\mathbf{f^{+}}=\mathbf{f_{d}^{+}}+\mathbf{f_{m}^{+}}$, meaning that the total focusing function $\mathbf{f^{+}}$ is decomposed into a direct arrival wave $\mathbf{f_{d}^{+}}$ and the following coda $\mathbf{f_{m}^{+}}$. Similar to the RM method, equation \ref{eq:UDRM_matrix_window} can be solved by the least-squares inversion (using, for example, the LSQR solver -- \cite{Paige1982}). Subsequently, the Green’s functions can be retrieved according to equation \ref{eq:UDRM_matrix}. However, in the presence of sparse receivers, the resulting focusing functions are affected by illumination artifacts (similar to those observed in classical seismic migration products). An alternative approach, later used in the numerical examples, involves re-casting the problem as a sparsity-promoting inversion with a suitable transform $\mathbf{S}$:
\begin{equation}
\label{eq:sparse}
\min_{\mathbf{z}_f}|| \mathbf{d}-\mathbf{G}\mathbf{S}^{H}\mathbf{z}_f||_{2}^{2}+\lambda || \mathbf{z}_f|| _{1}
,
\end{equation}
where $\mathbf{z}_f=[\mathbf{z}_{f^-}^T, \mathbf{z}_{f_m^+}^T]^T$ contains the sparse representation of the up-going and the coda of the down-going focusing functions, $\mathbf{d}$ and $\mathbf{G}$ identify the left-hand side vector and the right-hand side operator in equation \ref{eq:UDRM_matrix_window}, respectively. Finally $\lambda$ is a user-defined parameter to weight the role of sparsity in the inversion. Several optimization algorithms have been used in the literature to solve problems of such a kind for geophysical applications. Following \cite{Haindl2021} and \cite{Ravasi2023}, we use the Fast Iterative Shrinkage-Thresholding Algorithm (FISTA - \cite{Beck2009} ) solver coupled with a sliding linear Radon sparsifying transform. 

\section*{Numerical examples}

The proposed method is tested on two different synthetic datasets. 

\subsection{Syncline model}

\begin{figure}
  \centering
  \includegraphics[width=0.5\textwidth]{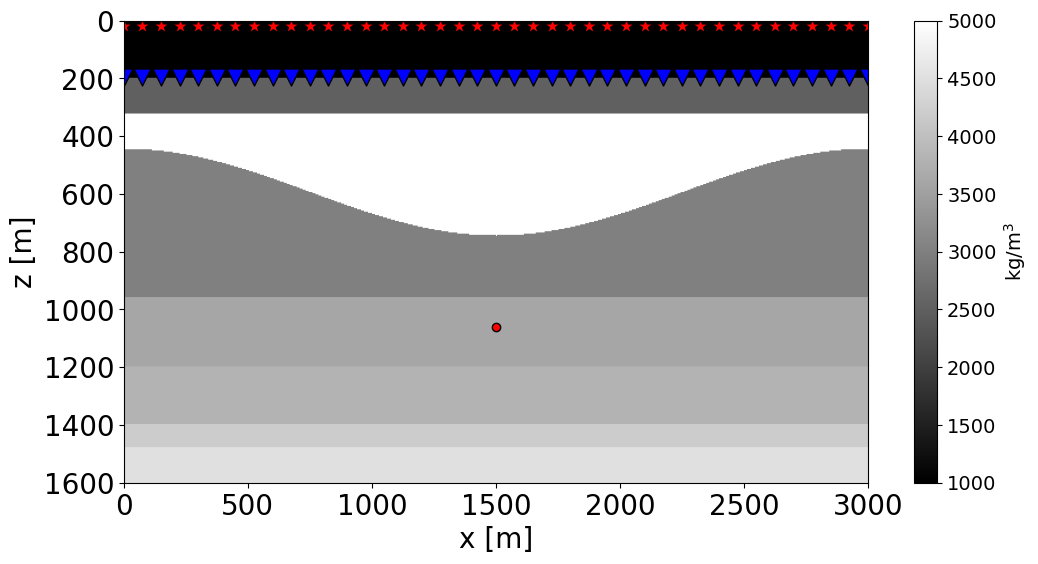}
  \caption{Syncline density model. Red stars identify sources, blue triangles represent receivers, and a red dot indicates the focal point used in the redatuming results.}
  \label{fig:geometry_simple}
\end{figure}

First, dual-source (i.e., monopole and dipole source), multi-component (pressure and particle velocity) data are modeled in a constant-velocity (2400 m/s), variable-density model (Figure \ref{fig:geometry_simple}) with free-surface effects included. In this case, we consider regular and densely sampled source and receiver arrays, which comprise of 201 sources at 20 m depth and 201 receivers at 196 m depth (i.e., along the seafloor). A focal point at x = 1500 m and z = 1060 m is considered in the redatuming results. 

\begin{figure*}
  \centering
  \includegraphics[width=0.87\textwidth]{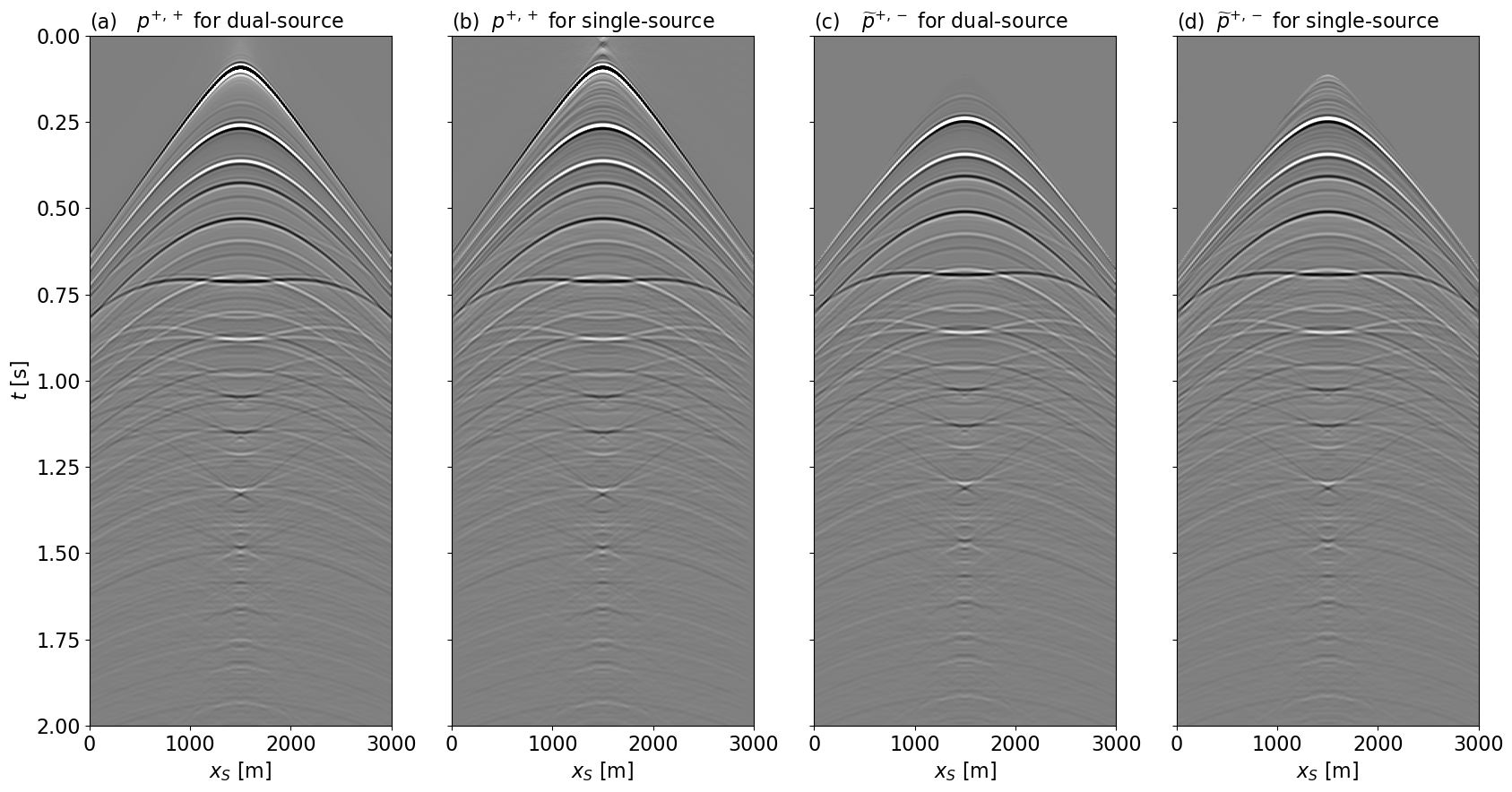}
  \caption{(a) Down-down-going wavefield computed using dual-source data, (b) down-down-going wavefield computed using single-source data, (c) down-up-going wavefield computed using dual-source data, (d) down-up-going wavefield computed using single-source data.}
  \label{fig:inputdata_simple}
\end{figure*}

To apply our method, we first remove the direct arrival from all available wavefields (i.e., pressure and particle velocity data from monopole and dipole sources). This wave is directly modelled in a constant-velocity, constant-density model. Next, the wavefields deprived of the direct arrival wave are separated along both the source and the receiver side. Wavefield separation is performed here by means of PZ summation~\citep{Wapenaar1998}. We first separate the wavefields from both the monopole and dipole sources into their up- and down-going components at the receiver side. The down-going components are further decomposed along the source side. However, dual sources are not always available in practice. In such situations, source-side deghosting can be employed to extract the source-side up- and down-going components of the data. The single-source data is still separated using PZ summation along the receiver side, while deghosting is applied to the receiver-side down-going component in order to accomplish wavefield separation at the source side. However, as deghosting may introduce some artifacts (Figure \ref{fig:inputdata_simple}), this can impact the quality of the subsequent redatuming and imaging results.

\begin{figure*}
  \centering
  \includegraphics[width=1\textwidth]{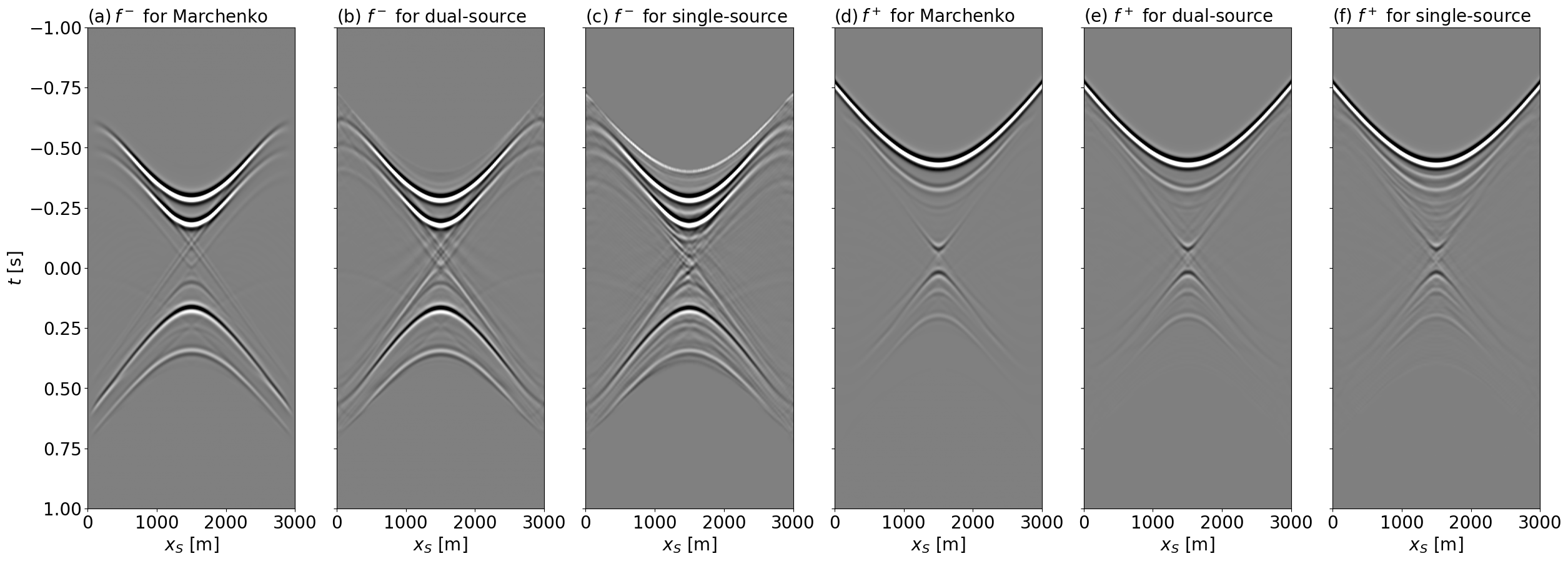}
  \caption{Up-going focusing function calculated by (a) the Marchenko method, (b) the UD-RM method with dual-source data, (c) the UD-RM method with single-source data. Down-going focusing function calculated by (d) the Marchenko method, (e) the UD-RM method with dual-source data, (f) the UD-RM method with single-source data.}
  \label{fig:FF_simple}
\end{figure*}

\begin{figure*}
  \centering
  \includegraphics[width=1\textwidth]{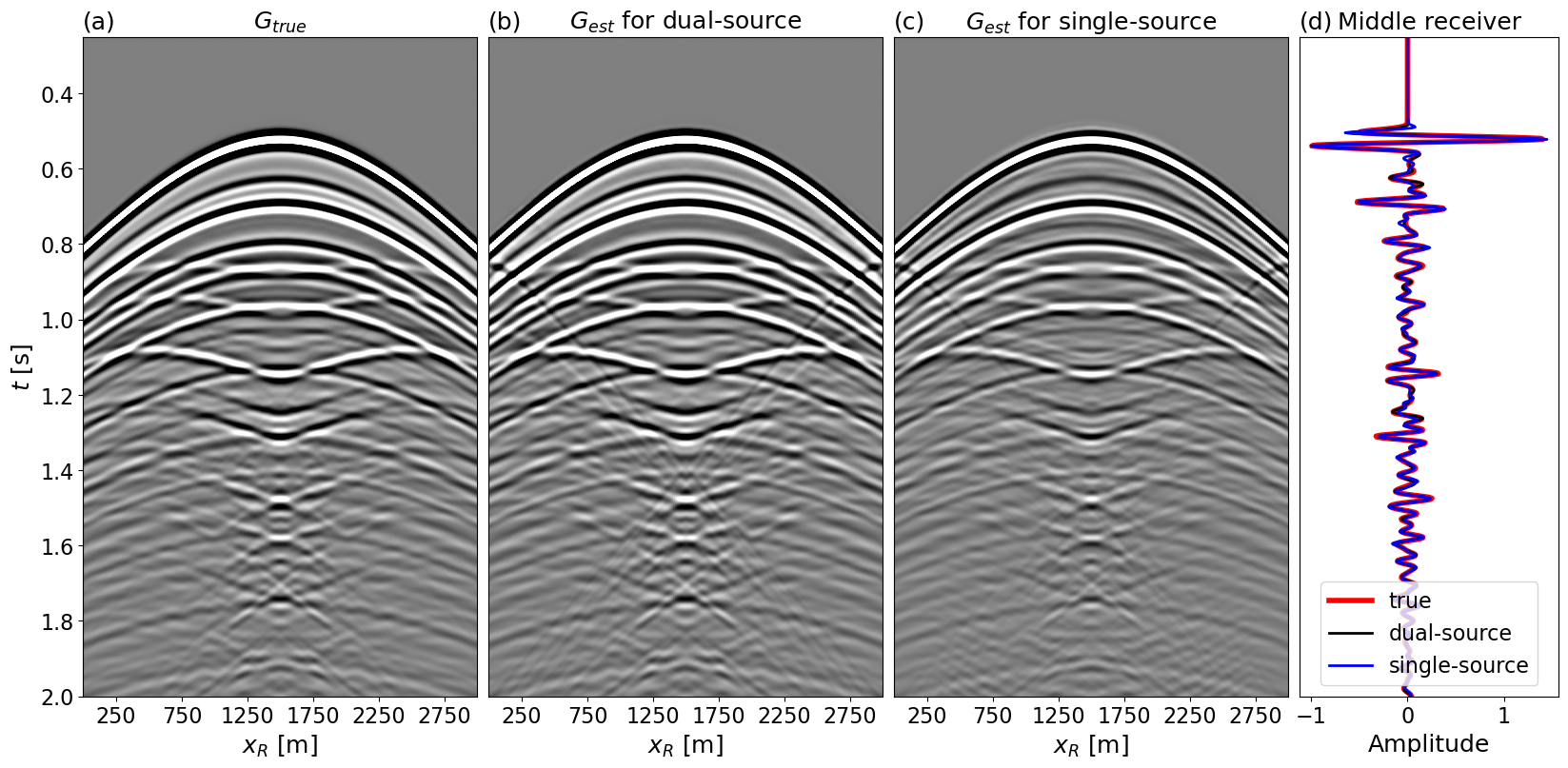}
  \caption{(a) True Green’s function obtained by finite-difference modeling, (b) Green’s function retrieved by the UD-RM method with dual-source data, and (c) Green’s function retrieved by the UD-RM method with single-source data. (d) Trace comparison between the true (red line), estimated with dual-source data (black line), and estimated with single-source data (blue line) Green’s functions in the middle receiver.}
  \label{fig:GF_simple}
\end{figure*}

The down-down-going wavefield $\partial _{z}p^{+,+}$ (down-going components at both the receiver side and source side) and down-up-going wavefield $\partial _{z}\widetilde{p}^{+,-}$ (down-going components at the receiver side and up-going components at source side) calculated by these two approaches are now used to solve equation \ref{eq:UDRM_matrix_window} using the LSQR solver~\citep{Paige1982}, thereby yielding the up- and down-going focusing functions shown in Figure \ref{fig:FF_simple}. To validate the accuracy of the UD-RM method, another dataset is modeled by placing receivers at the same locations as sources and excluding the influence of the free-surface. Subsequently, the focusing functions are computed using the Marchenko method from this new dataset. As shown in the Theory section, the focusing functions calculated by Marchenko method and UD-RM method must be identical. From Figure \ref{fig:FF_simple}, the focusing functions from the UD-RM and Marchenko methods exhibit a substantial degree of similarity, with only minor discrepancies observed in the results of the UD-RM method using dual-source data. This may be caused by the different convergence of the LSQR solver. The focusing functions computed by the UD-RM method with single-source data present more artifacts, likely due to the deghosting process (Figures \ref{fig:FF_simple}c and \ref{fig:FF_simple}f). The total estimated Green’s functions (Figures \ref{fig:GF_simple}b and \ref{fig:GF_simple}c) are determined by evaluating equation \ref{eq:UDRM_matrix} with the retrieved focusing functions and compared with the true Green’s function (Figure \ref{fig:GF_simple}a) modeled via finite-difference. The traces, located in the middle of the receiver array, as depicted in Figure \ref{fig:GF_simple}d, exhibit an almost perfect match, validating the effectiveness of the proposed method.

\begin{figure*}
  \centering
  \includegraphics[width=0.76\textwidth]{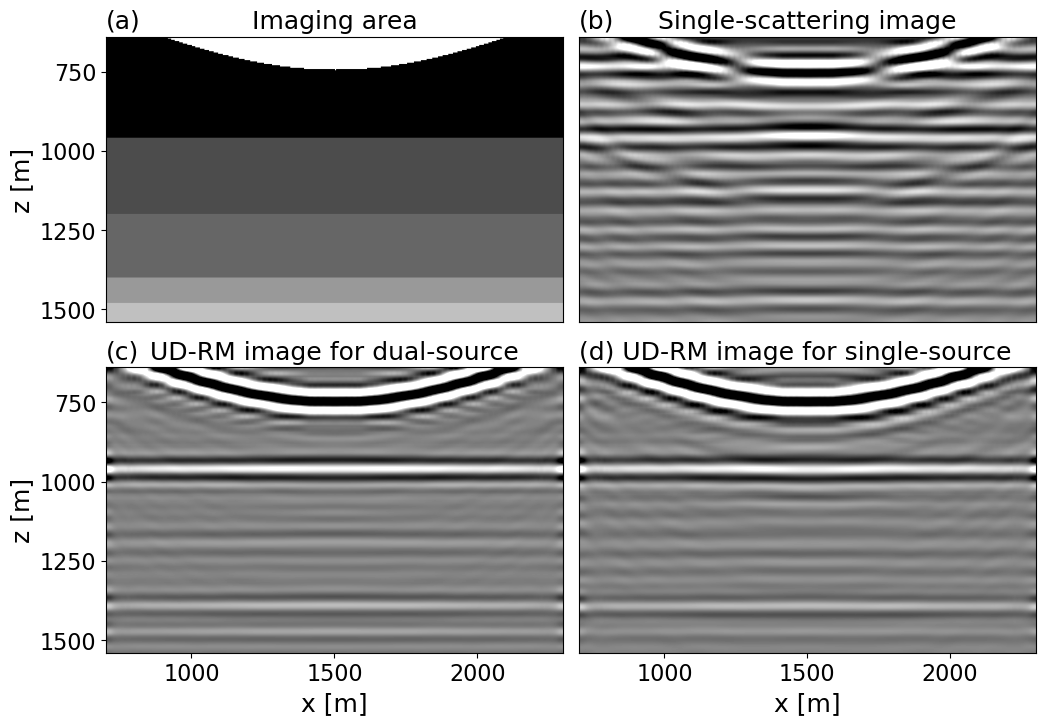}
  \caption{(a) The target area used for imaging, (b) image computed by single-scattering imaging with dual-source data, (c) image computed by the UD-RM method with dual-source data, (d) image computed by the UD-RM method with single-source data.}
  \label{fig:imaging_simple}
\end{figure*}

\begin{figure*}
  \centering
  \includegraphics[width=0.92\textwidth]{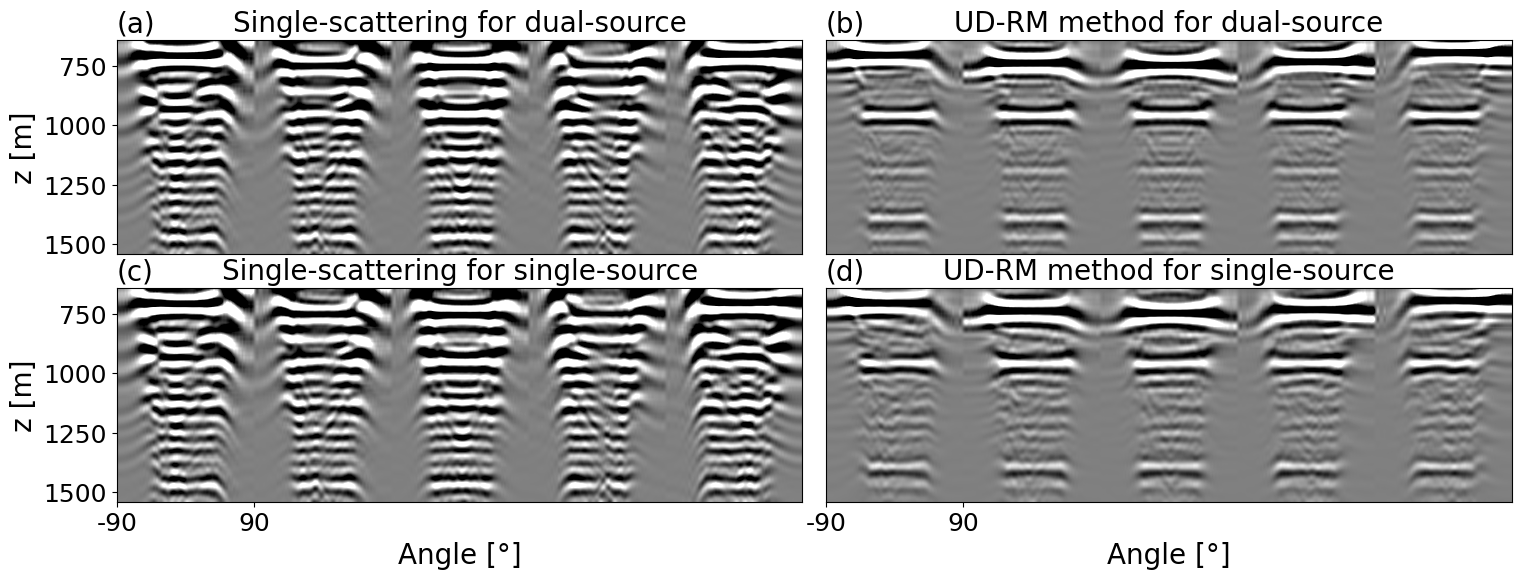}
  \caption{Angle gathers for (a) single-scattering with dual-source data, (b) the UD-RM method with dual-source data, (c) single-scattering with single-source data, (d) the UD-RM method with single-source data.}
  \label{fig:anglegather_simple}
\end{figure*}

Subsequently, Green's functions are computed in a grid (from x = 700 m to 2300 m and from z = 640 m to 1540 m, as shown in Figure \ref{fig:imaging_simple}a) with a vertical and horizontal spacing of 20 m. These Green's functions are then used to perform imaging as described in \cite{Wapenaar2017}. Single-scattering, source-receiver imaging, which is equivalent to using Green’s functions obtained from the first iteration of the UD-RM iterative scheme, is also carried out for comparison. Since the Green's function obtained from the single-source data produce worse imaging results, here we only show the single-scattering image calculated using dual-source data (Figure \ref{fig:imaging_simple}b). Due to the strong surface-related multiples, the subsurface structure cannot be discerned from the single-scattering image, proving that single-scattering imaging cannot be applied on unprocessed data. On the contrary, the image computed using the wavefields from the UD-RM redatuming method with dual-source data (Figure \ref{fig:imaging_simple}c) presents a good match with the true density model. The image produced by the UD-RM method with single-source data (Figure \ref{fig:imaging_simple}d) is also of good quality, with most multiple-related artifacts being effectively removed. In order to further investigate the imaging capabilities of our method (without fully to leveraging the stacking power of migration), we randomly select five horizontal locations (at index 21, 31, 41, 51, and 61) from the redatumed common-virtual source gathers and calculate their true-amplitude subsurface angle gathers~\citep{de1990, Ordoñez2014}. Comparing them with those produced by single scattering imaging for the two different input datasets (Figures \ref{fig:anglegather_simple}a and \ref{fig:anglegather_simple}c) reveals that the angle gathers of UD-RM imaging (Figures \ref{fig:anglegather_simple}b and \ref{fig:anglegather_simple}d) are cleaner and more clearly show the angle dependent responses from key reflectors in the model.

\begin{figure}
  \centering
  \includegraphics[width=0.5\textwidth]{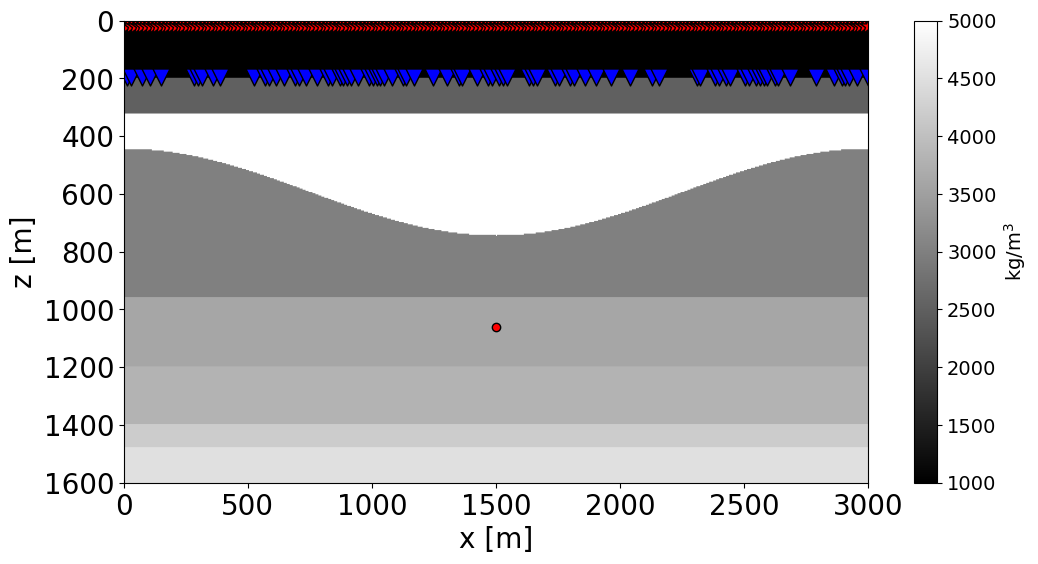}
  \caption{Syncline density model with sparse receiver acquisition. Keys as in Figure \ref{fig:geometry_simple}.}
  \label{fig:geometry_simple_sparse}
\end{figure}

\begin{figure*}
  \centering
  \includegraphics[width=1\textwidth]{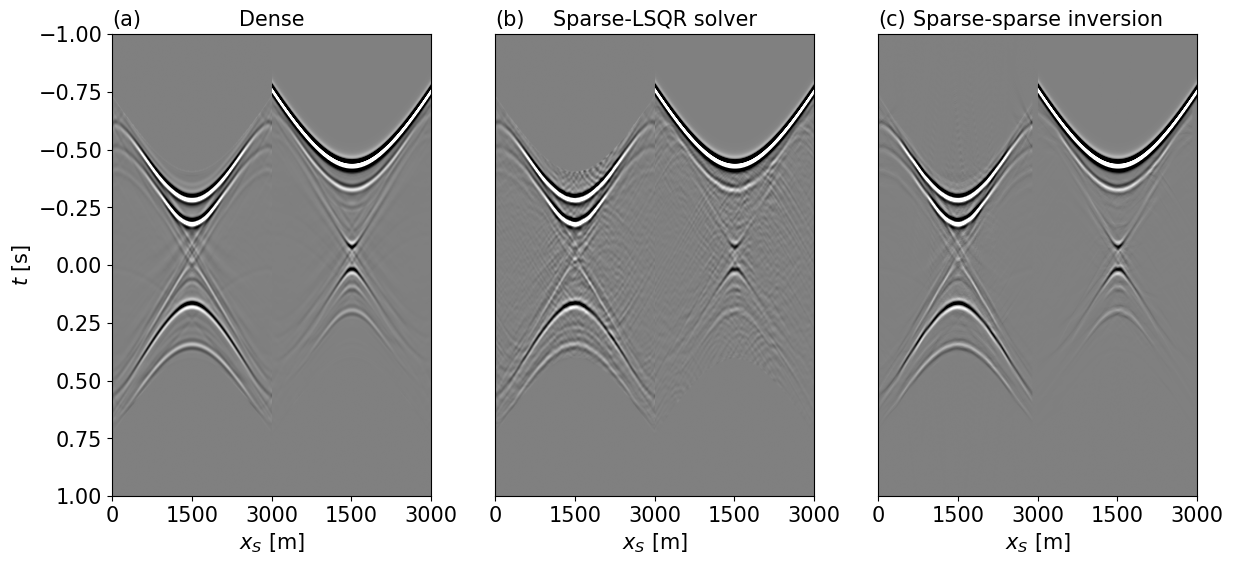}
  \caption{Focusing functions obtained from (a) the dense receiver array, (b) the sparse receiver array calculated by the LSQR solver, (c) the sparse receiver array calculated by sparsity-promoting inversion.}
  \label{fig:FF_simple_sparse}
\end{figure*}

\begin{figure*}
  \centering
  \includegraphics[width=1\textwidth]{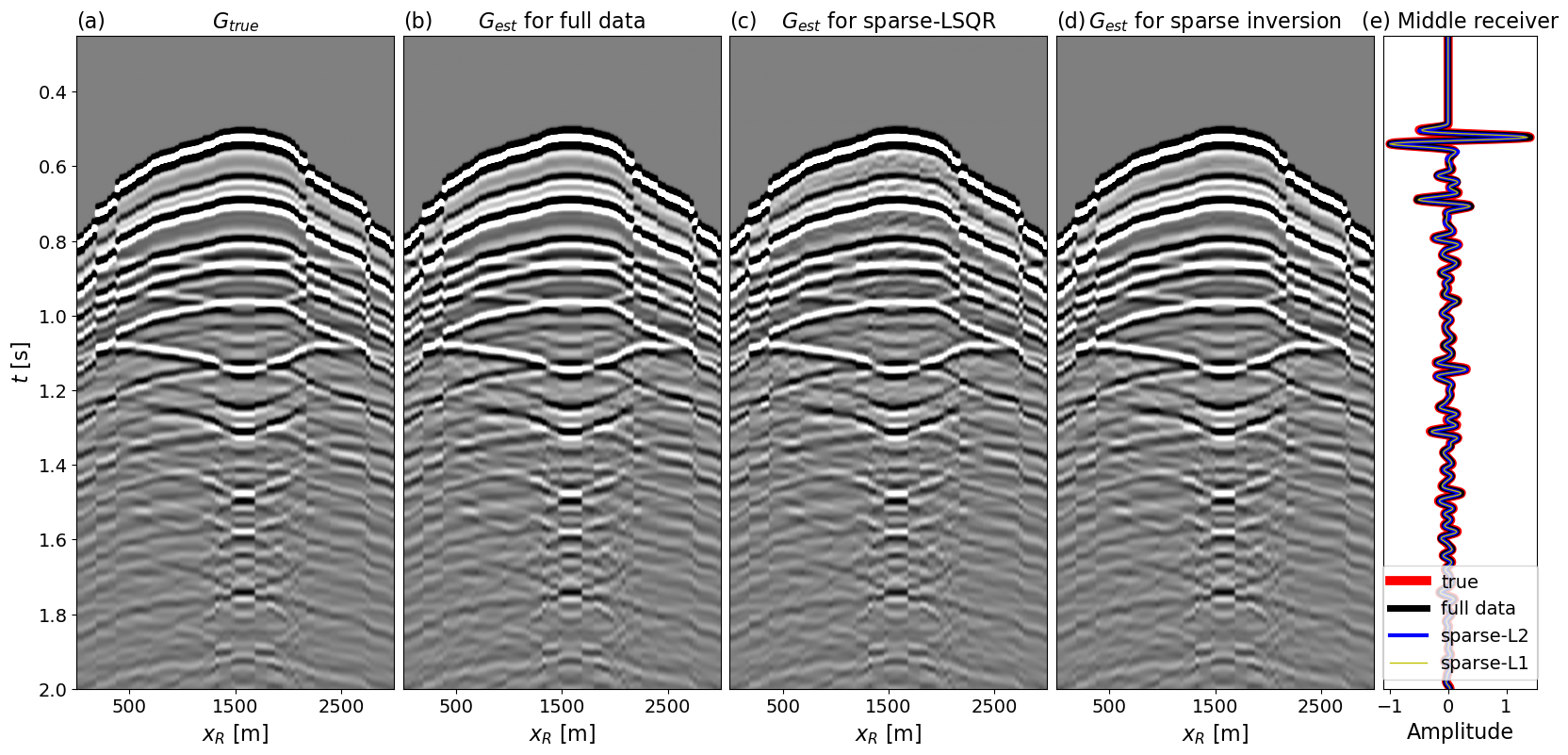}
  \caption{(a) True Green’s function computed via finite-difference modeling, (b) Green’s function obtained by solving the UD-RM equations with a dense receiver array, (c) Green’s function obtained from the LSQR solution in the presence of a sparse receiver array, and (d) Green’s function obtained by sparsity-promoting inversion for the sparse receiver array. (e) Trace comparison in the middle receiver between the true (red line), estimated with full data (black line), estimated by LSQR solution with sparse data (blue line), and estimated by sparsity-promoting inversion with sparse data (yellow line).}
  \label{fig:GF_simple_sparse}
\end{figure*}

We now proceed to evaluate the ability of the UD-RM method to handle sparse receiver arrays by randomly selecting 40 percent of the total receivers (Figure \ref{fig:geometry_simple_sparse}). As previously discussed in the Theory section, removing receivers renders equation \ref{eq:UDRM_matrix_window} to be a heavily under-determined inverse problem. Two strategies are assessed to estimate the focusing functions in such a scenario: first, we solve these equations with 30 iterations of the LSQR solver (this is the same strategy used in the previous scenario) leading to the focusing functions in Figure \ref{fig:FF_simple_sparse}b. When comparing these focusing functions with the ones retrieved from the dense receiver array (Figure \ref{fig:FF_simple_sparse}a), we can conclude that they exhibit some artifacts, which further impact the quality of the retrived Green's function (Figure \ref{fig:GF_simple_sparse}c). Second, in an attempt to further enhance the quality of the focusing functions, we solve the problem in equation \ref{eq:sparse}. In this case, the FISTA solver is run for a total of 200 iterations, yielding focusing functions (Figure \ref{fig:FF_simple_sparse}c) that are cleaner and better resemble those obtained from the dense receiver array (Figure \ref{fig:FF_simple_sparse}a). To ease visual comparison, only the traces at the available receivers are shown for the resulting Green's functions (Figure \ref{fig:GF_simple_sparse}). We can conclude that with the help of a sliding linear Radon transform as a sparse constraint, the obtained Green's function (Figure \ref{fig:GF_simple_sparse}d) exhibit a better agreement with the true counterpart; most of the illumination artifacts are suppressed (see for example the earlier arrivals around 0.6s), as further substantiated by the comparative analysis of the middle trace (Figure \ref{fig:GF_simple_sparse}e). However, it is worth pointing out that since sparse inversion algorithms are generally slower in terms of convergence (i.e., FISTA requires 200 iterations to converge to a satisfactory solution in this example), this approach is more time-consuming than its least-squares counterpart.

\begin{figure*}
  \centering
  \includegraphics[width=0.76\textwidth]{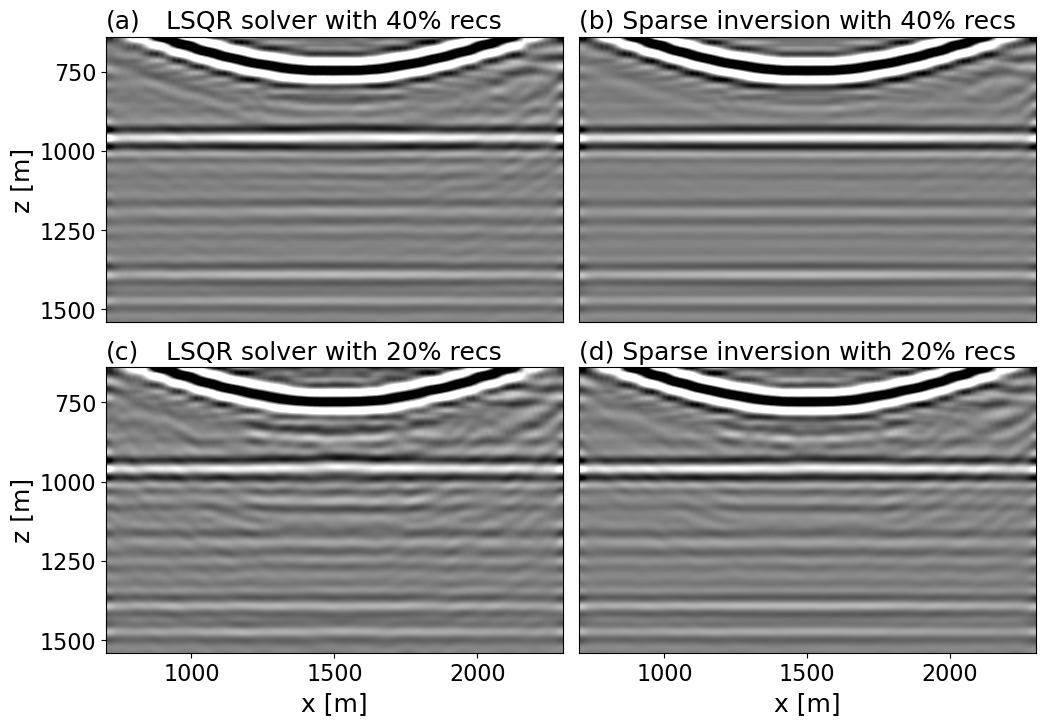}
  \caption{Images computed using the Green's functions estimated by (a) least-squares inversion for 40\% of receivers, (b) sparse inversion for 40\% of receivers, (c) least-squares inversion for 20\% of receivers, and (d) sparse inversion for 20\% of receivers.}
  \label{fig:imaging_simple_sparse}
\end{figure*}

\begin{figure*}
  \centering
  \includegraphics[width=0.92\textwidth]{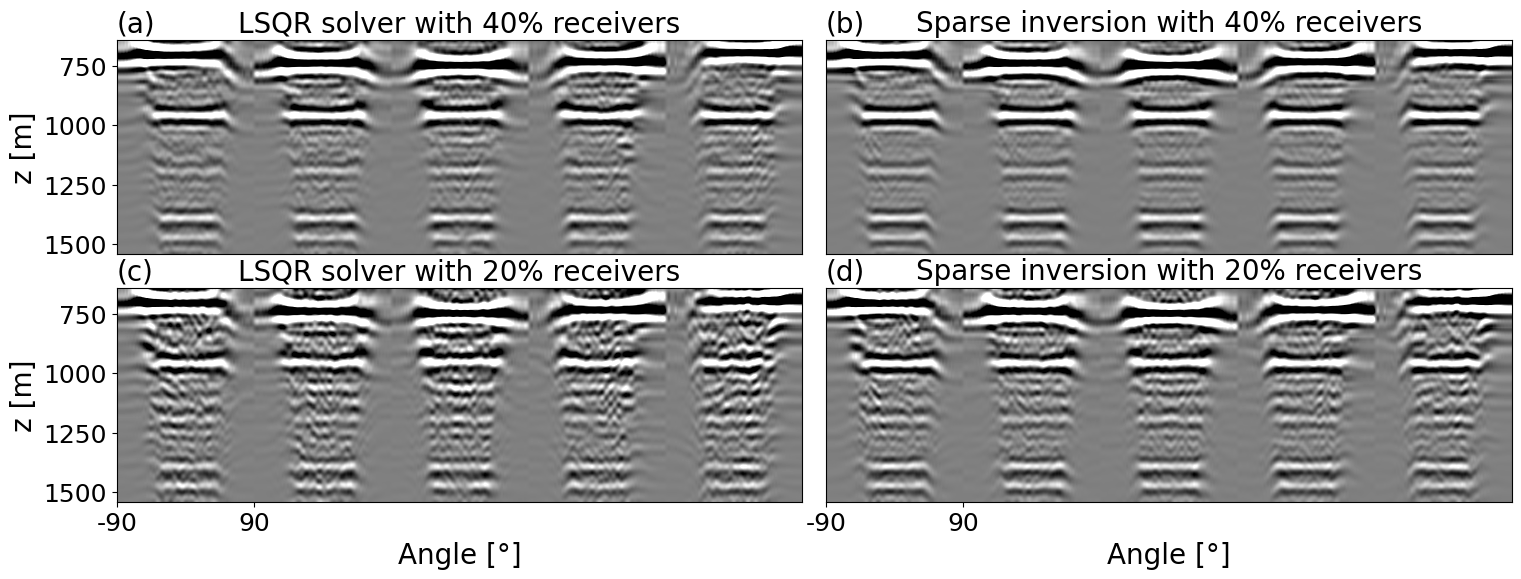}
  \caption{Angle gathers computed by the UD-RM method with (a) least-squares inversion for 40\% of receivers, (b) sparse inversion for 40\% of receivers, (c) least-squares inversion for 20\% of receivers, (d) sparse inversion for 20\% of receivers.}
  \label{fig:anglegather_simple_sparse}
\end{figure*}

Imaging is finally carried out in the same area used in the previous examples (Figure \ref{fig:imaging_simple}a) with the same grid spacing (20 m) using the Green's functions retrieved by the UD-RM method based on least-squares and sparse inversion methods. As shown in Figures \ref{fig:imaging_simple_sparse}a and \ref{fig:imaging_simple_sparse}b, the proposed method can deal well with sparse receiver arrays; this is further verified by the angle gathers (Figures \ref{fig:anglegather_simple_sparse}a and \ref{fig:anglegather_simple_sparse}b). When comparing the results from the least-squares and sparse inversion methods, we can conclude that the noise presents in the Green's functions for the least-squares inversion does not visibly appear in the resulting image. Nevertheless, the angle gathers reveal the benefit of applying sparse inversion when the outputs of the UD-RM method are used in subsequent reservoir characterization studies. Next, we reduce the percentage of available receivers to 20 percent of the original one. The resulting images are shown in Figures \ref{fig:imaging_simple_sparse}c and \ref{fig:imaging_simple_sparse}d, and the corresponding angle gathers in Figures \ref{fig:anglegather_simple_sparse}c and \ref{fig:anglegather_simple_sparse}d. Even only with 20 percent of receivers, the imaging results based on the Green's functions obtained by means of sparse inversion are still satisfactory. On the other hand, those produced by Green's functions obtained from the LSQR solution contain larger artifacts that affect the identification of the key geological boundaries, especially in the middle part of the imaging area. From these results, we can conclude that whilst reducing the number of available receivers can affect, not surprisingly, the quality of redatuming and imaging, the results obtained with the UD-RM method prove its effectiveness in handling realistic sparse seabed geometries. It is worth noting that all of the imaging results with sparse receiver geometries have been carried out using dual-source data. In the presence of single-source data, we would expect the resulting imaging products to be of slightly lower quality, as shown above for the dense receiver scenario.
 
\subsection{SEG/EAGE Overthrust model}
\begin{figure*}
  \centering
  \includegraphics[width=0.98\textwidth]{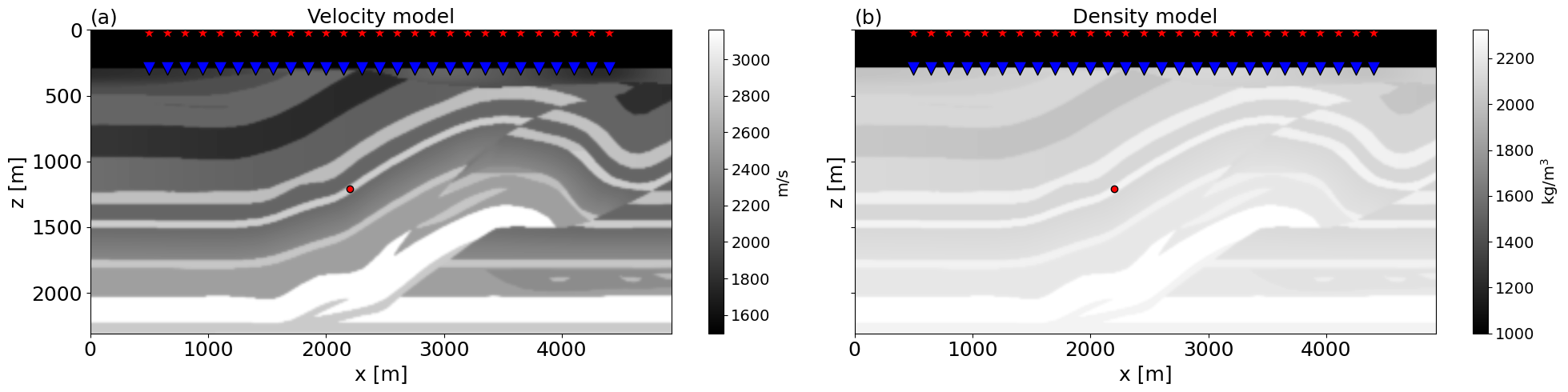}
  \caption{Modified Overthrust model. (a) Velocity and (b) density models.}
  \label{fig:geometry_overthrust}
\end{figure*}

In order to verify the effectiveness of the proposed method in a more realistic geological setting, we consider a variable-density, variable-velocity model (Figure \ref{fig:geometry_overthrust}). The velocity model corresponds to a 2D slice of the SEG/EAGE Overthrust model~\citep{Aminzadeh1997}, while the density model is constructed using Gardner's equation~\citep{Gardner1974}. A 288 m thick water column with velocity of 1500 m/s and density of 1000 kg/m\textsuperscript{3} is added on top of the original model to simulate a seabed seismic acquisition scenario. 401 sources are placed at a depth of 20 m and 401 receivers are placed along the seafloor, both of which are uniformly distributed from x = 500 m to x = 4500 m. A focus point is selected at x = 2200 m and z = 1206 m. 

\begin{figure*}
  \centering
  \includegraphics[width=0.98\textwidth]{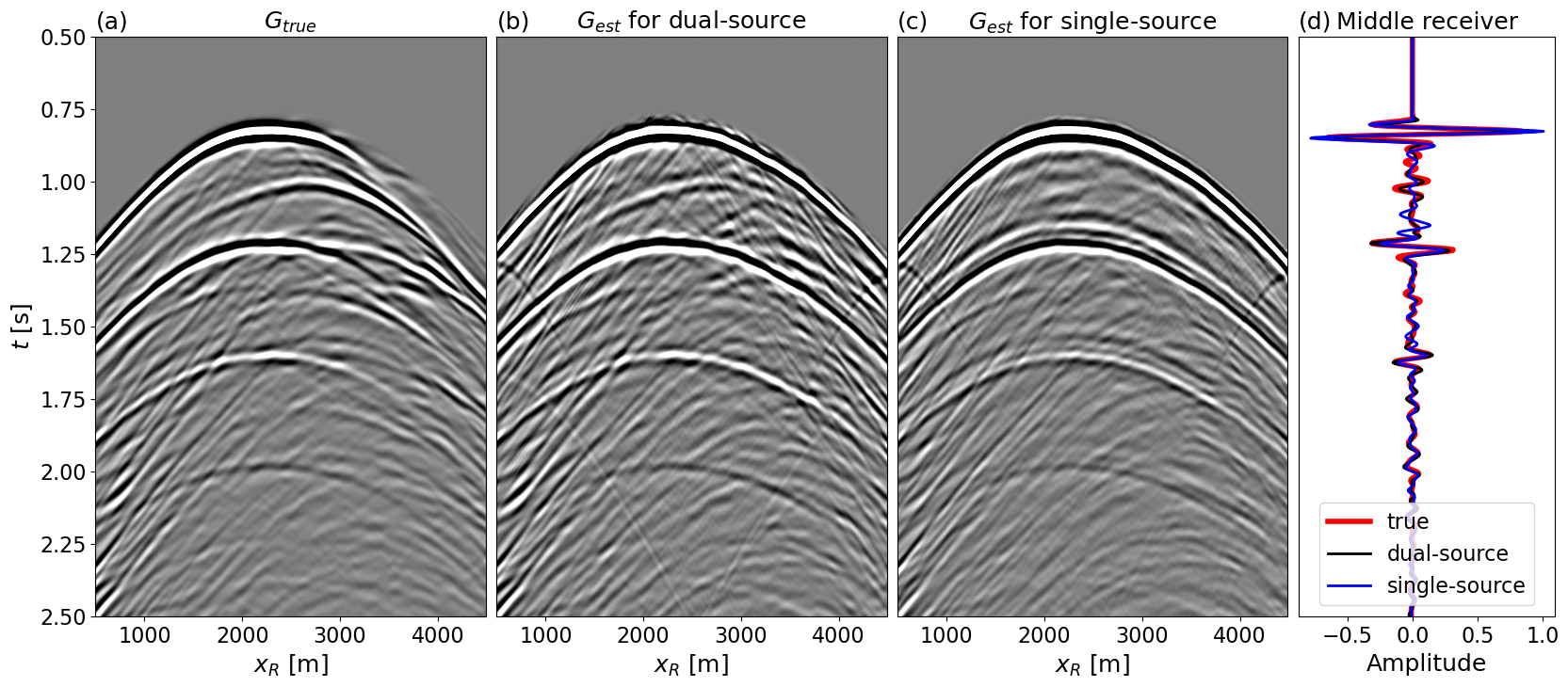}
  \caption{(a) True Green’s function obtained by finite-difference modeling for the Overthrust model, (b) Green’s function retrieved by the UD-RM method with dual-source data, and (c) Green’s function retrieved by the UD-RM method with single-source data. (d) Trace comparison between the true (red line), estimated with dual-source data (black line), and estimated with single-source data (blue line) Green’s functions in the middle receiver.}
  \label{fig:GF_overthrust}
\end{figure*}

Similar to the previous example, we process dual-source data and single-source data using PZ-summation and deghosting, respectively. In both cases, we perform least-squares inversion to estimate the focusing functions and subsequently calculate the corresponding Green's functions. When comparing the estimated Green's functions with the true one calculated by finite-difference modeling (Figure \ref{fig:GF_overthrust}a), we notice that small artifacts are present in the estimated Green's function of the dual-source data (Figure \ref{fig:GF_overthrust}b). We conjecture that the main cause of these artifacts may be the imperfect cancellation of the direct wave (due to velocity variations along the seafloor, which does not allow us to perfectly model and subtract the direct wave from the recorded data). Moreover, deghosting generates additional artifacts in the down-going wavefield, and the Green's function obtained from the single-source data contains additional artifacts (Figure \ref{fig:GF_overthrust}c). The traces in the middle receiver array are displayed in Figure \ref{fig:GF_overthrust}d; though the estimated Green's functions and the true one are not perfectly fitted in some locations (especially for the single-source data), the main events are accurately matched in both cases. 

\begin{figure*}
  \centering
  \includegraphics[width=0.98\textwidth]{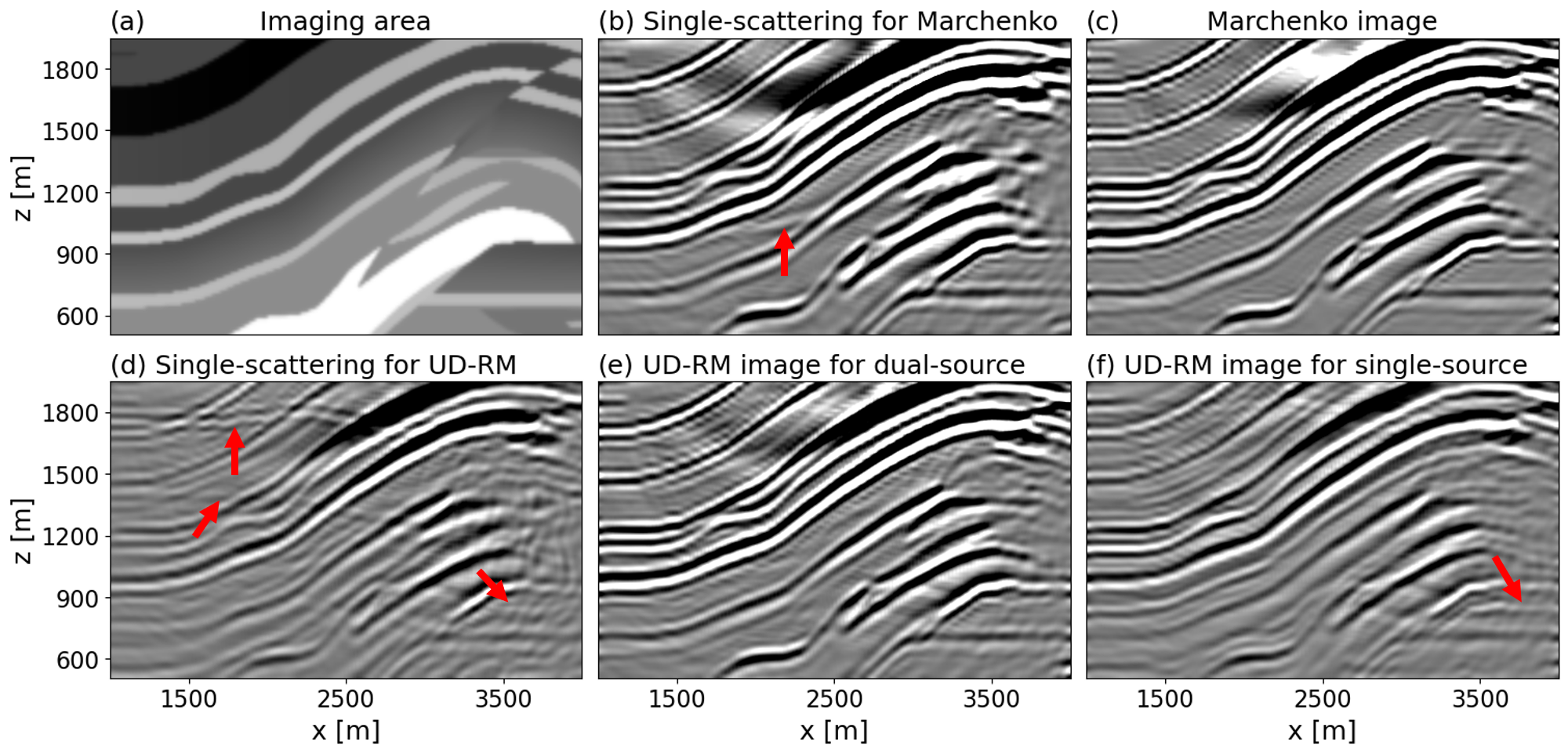}
  \caption{(a) The target area used for imaging, (b) image computed by the single-scattering imaging without the free-surface effects, (c) image computed by the original Marchenko method without the free-surface effects, (d) image computed by the single-scattering imaging with surface-related multiples, (e) image computed by the UD-RM imaging using dual-source data, (f) image computed by the UD-RM imaging using single-source data.}
  \label{fig:imaging_overthrust_arrow}
\end{figure*}

\begin{figure*}
  \centering
  \includegraphics[width=0.92\textwidth]{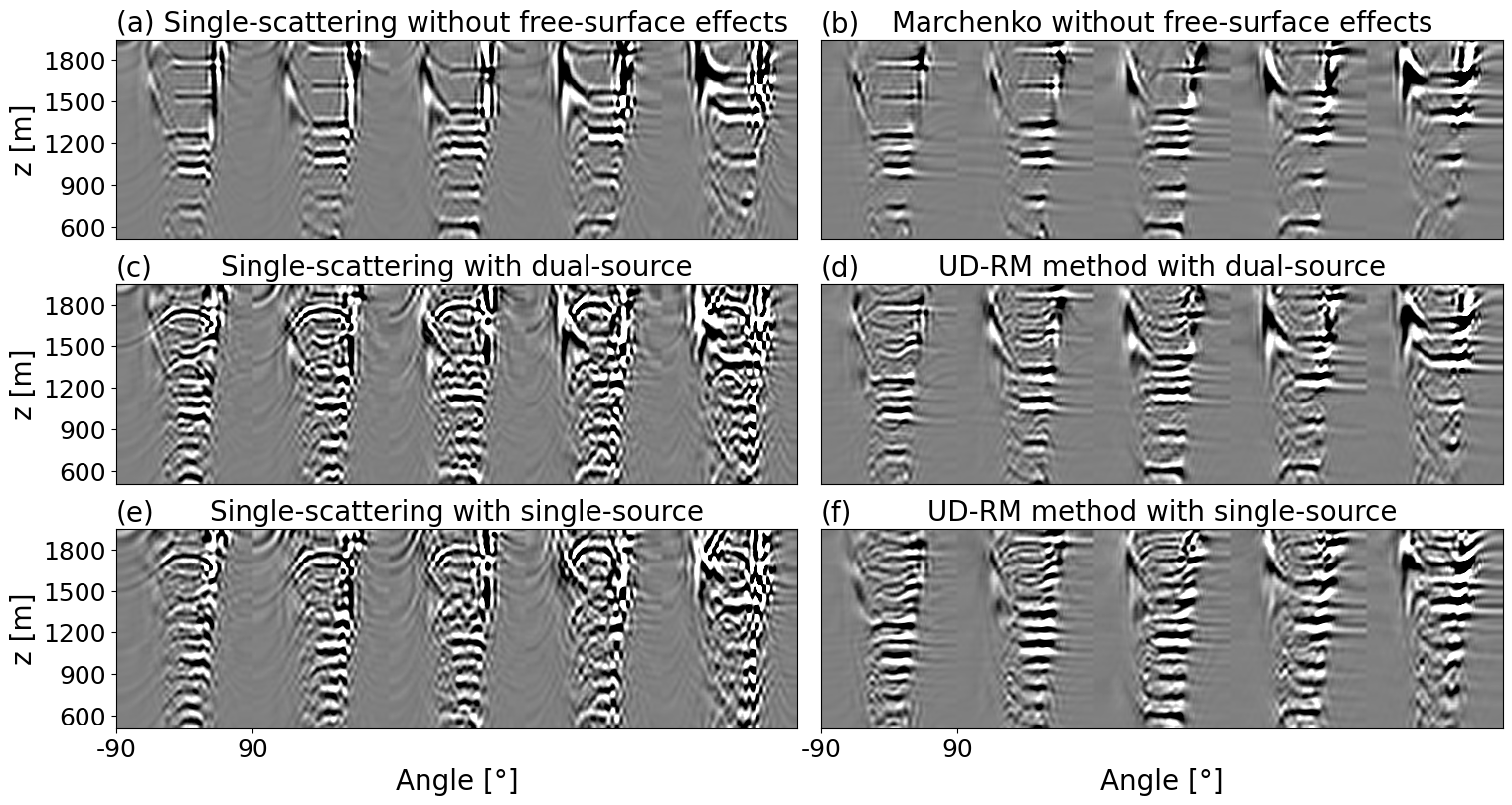}
  \caption{Angle gathers for (a) single-scattering without the free-surface effects, (b) the Marchenko method without the free-surface effects, (c) single-scattering with dual-source data, (d) the UD-RM method with dual-source data, (e) single-scattering with single-source data, (f) the UD-RM method with single-source data.}
  \label{fig:anglegather_overthrust}
\end{figure*}

Next, imaging is carried out within a grid with a vertical and horizontal spacing of 24 m (Figure \ref{fig:imaging_overthrust_arrow}a). Another dataset without free-surface effects and with both sources and receivers at a depth of 20 m is modelled to carry out Marchenko imaging (Figure \ref{fig:imaging_overthrust_arrow}c); as expected, this benchmark image shows good agreement with the true model. Single-scattering imaging is also performed using data without free-surface multiples (Figure \ref{fig:imaging_overthrust_arrow}b). Due to the weaker contribution of internal multiples, only small differences can be observed between the single-scattering image and the Marchenko image (as indicated by the red arrow in Figure \ref{fig:imaging_overthrust_arrow}b). Meanwhile, single-scattering imaging using the data with surface-related multiples is strongly affected by artifacts associated to surface-related multiples (as marked by the red arrows in Figure \ref{fig:imaging_overthrust_arrow}d). On the other hand, artifacts from both the surface-related multiples and internal multiples are effectively suppressed in the image produced from the UD-RM method (Figure \ref{fig:imaging_overthrust_arrow}e), proving the ability of UD-RM to handle complex subsurface models. Finally, as shown in Figure \ref{fig:imaging_overthrust_arrow}f, when source deghosting is applied to separate the data on the source-side, the quality of the resulting image is slightly compromised. A similar conclusion can be drawn by looking at the angle gathers: single scattering imaging (Figure \ref{fig:anglegather_overthrust}a) produces cleaner angle gathers for data that do not contain surface-related multiples compared to data with free-surface effects (Figures \ref{fig:anglegather_overthrust}c and \ref{fig:anglegather_overthrust}e). The angle gathers produced by the Marchenko method are instead successfully deprived of any artefact associated with internal multiples in the data (Figure \ref{fig:anglegather_overthrust}b). The UD-RM method provides angle gathers of comparable quality, revealing once again its ability to dealing with both surface-related and internal multiples (Figures \ref{fig:anglegather_overthrust}d and \ref{fig:anglegather_overthrust}f). Similar to the imaging results, the angle gathers from single source-data are slightly noisier with visible residual artefacts.

\begin{figure*}
  \centering
  \includegraphics[width=1\textwidth]{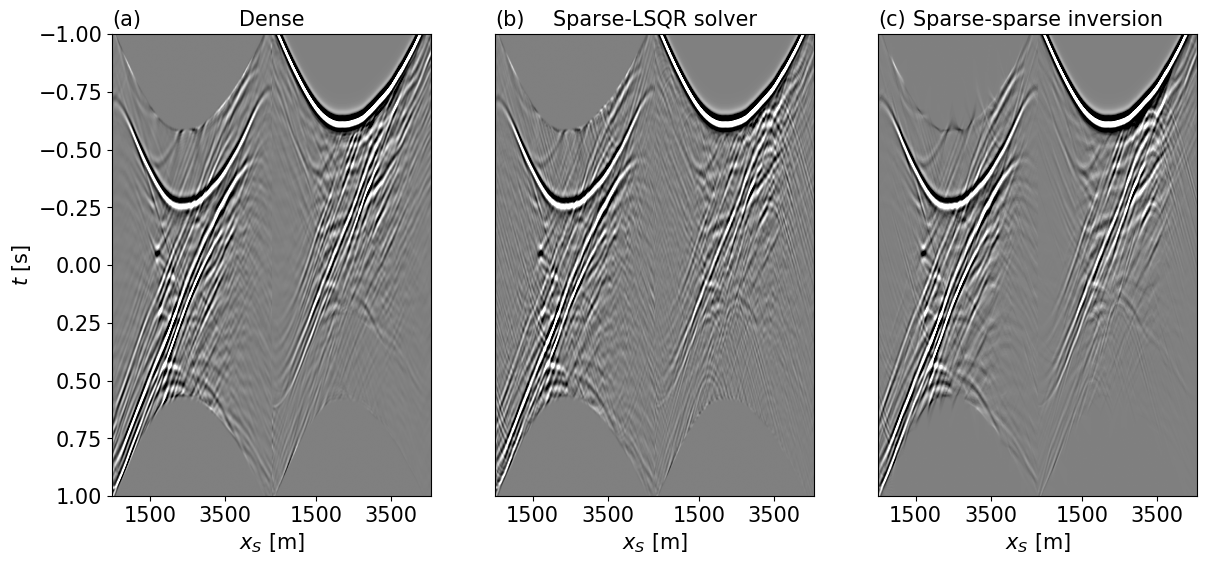}
  \caption{Focusing functions for the Overthrust model obtained from (a) the dense receiver array, (b) the sparse receiver array calculated by least-squares inversion, (c) the sparse receiver array calculated by sparsity-promoting solution .}
  \label{fig:FF_overthrust_sparse}
\end{figure*}

\begin{figure*}
  \centering
  \includegraphics[width=1\textwidth]{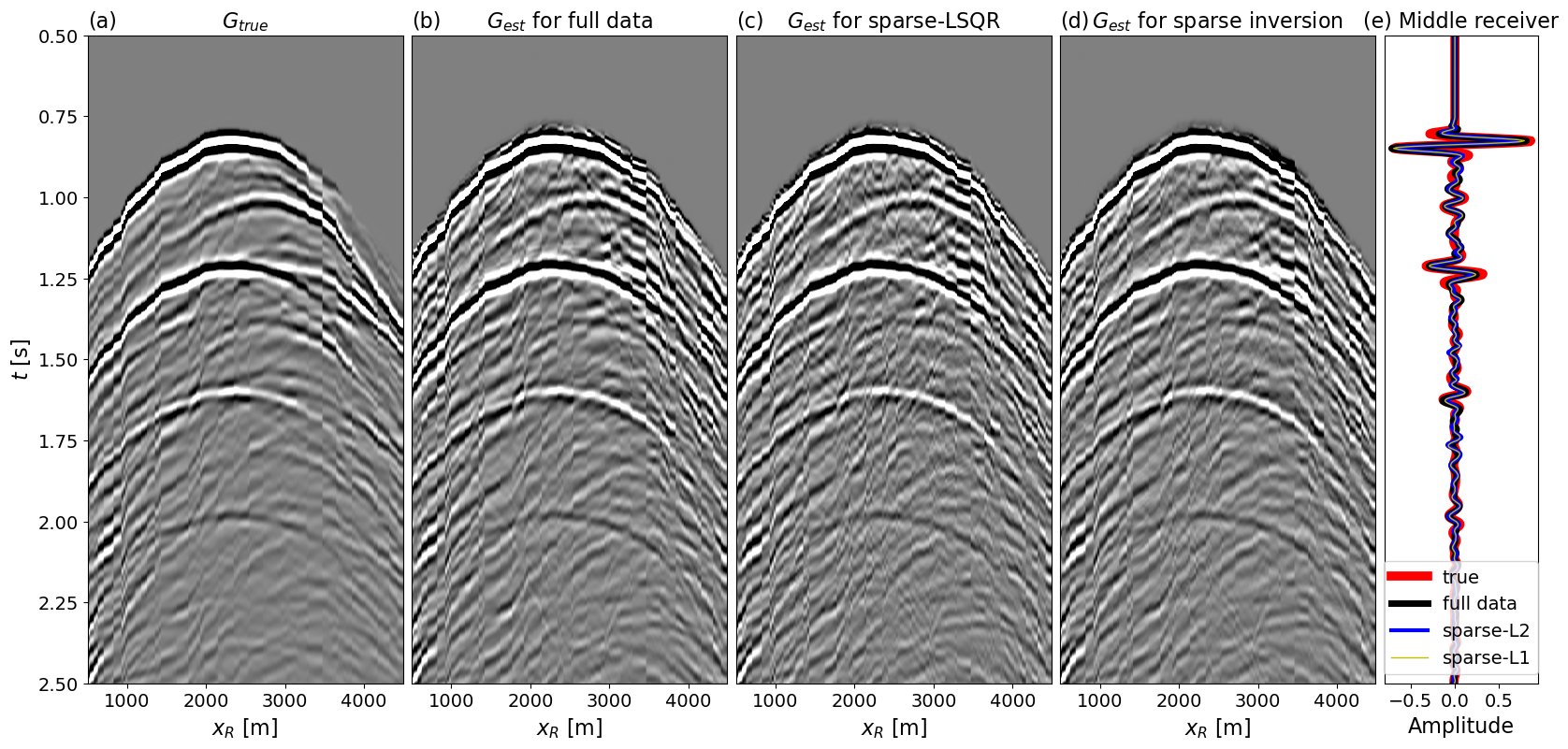}
  \caption{(a) True Green’s function obtained by finite-difference modeling, (b) Green’s function obtained from the dense receiver array, (c) Green’s function obtained by least-squares inversion from the sparse receiver array, and (d) Green’s function obtained by sparsity-promoting solution from the sparse receiver array. (e) Trace comparison between the true (red line), estimated with full data (black line), estimated by least-squares inversion with sparse data (blue line), and estimated by sparsity-promoting solution with sparse data (yellow line) Green’s functions in the middle receiver.}
  \label{fig:GF_overthrust_sparse}
\end{figure*}

\begin{figure*}
  \centering
  \includegraphics[width=0.82\textwidth]{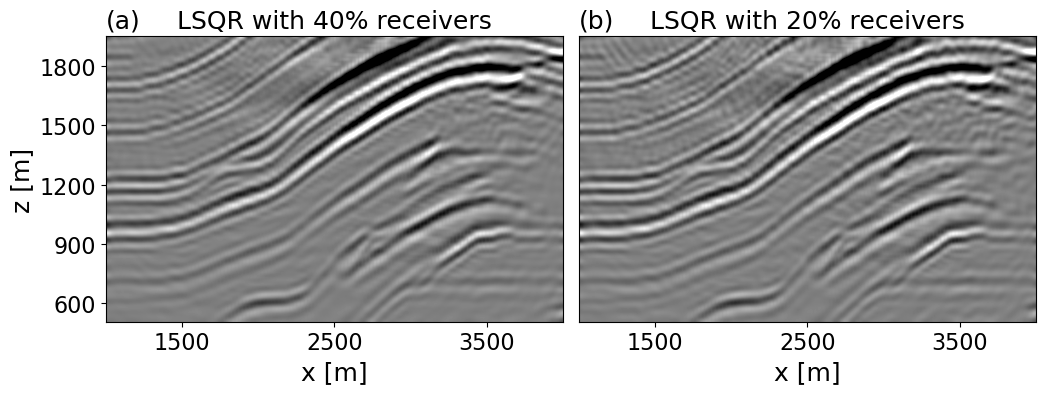}
  \caption{Image of the Overthrust model obtained from least-squares inversion with (a) 40\% and (b) 20\% of receivers.}
  \label{fig:imaging_overthrust_sparsel2}
\end{figure*}

\begin{figure*}
  \centering
  \includegraphics[width=0.92\textwidth]{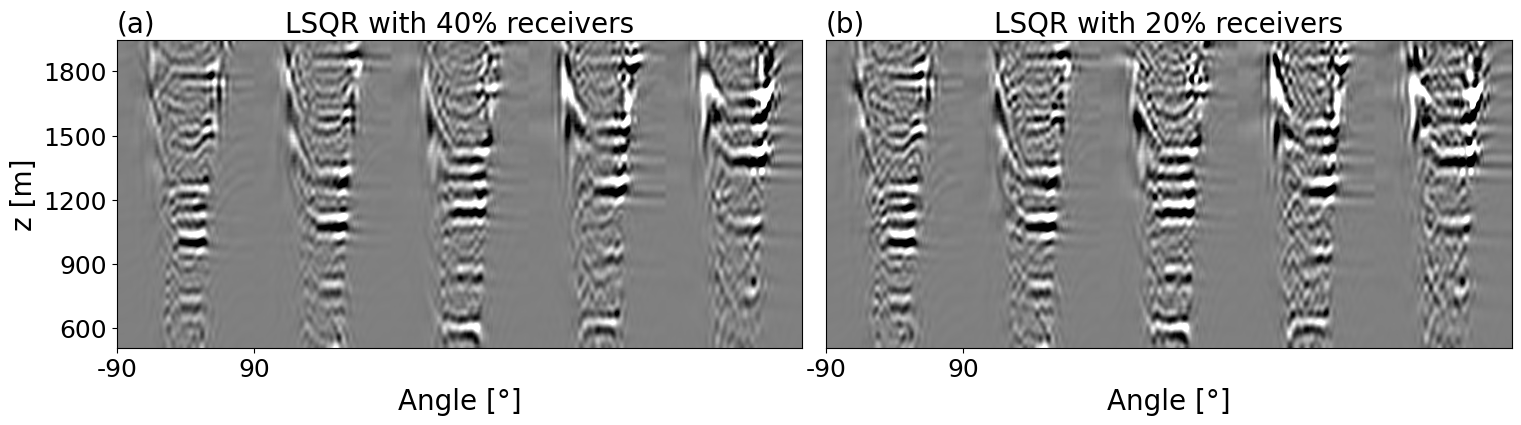}
  \caption{Angle gathers of the Overthrust model obtained from least-squares inversion with (a) 40\% and (b) 20\% of receivers.}
  \label{fig:anglegather_overthrust_sparsel2}
\end{figure*}

Next, we randomly select 40 percent of the receivers. The same methods used in the previous example are used to calculate the focusing functions (Figures \ref{fig:FF_overthrust_sparse}b and \ref{fig:FF_overthrust_sparse}c). Comparing these focusing functions with those obtained from the dense receiver array (Figure \ref{fig:FF_overthrust_sparse}a) reveals that sparse inversion is more capable at suppressing illumination-related artefacts than its least-squares counterpart. As far as the resulting Green's functions (Figure \ref{fig:GF_overthrust_sparse}) are concerned, both inversion results match well with the true wavefield and the Green's function obtained by sparse inversion is cleaner. However, due to the higher computational cost of the sparse inversion, in this example we use least-squares inversion to construct images in the same imaging area used for the case of dense receiver array (Figure \ref{fig:imaging_overthrust_sparsel2}). Figure \ref{fig:imaging_overthrust_sparsel2}a reveals that performing imaging with the redatumed wavefields from least-squares inversion can provide a satisfactory representation of the subsurface main structures with good suppression of multiple-related artefacts and limited noise from illumination-based artefacts, which is comparable to the image created from data with a dense receiver array (Figure \ref{fig:imaging_overthrust_arrow}e). Figure \ref{fig:imaging_overthrust_sparsel2}b shows the imaging result when the percentage of selected receivers is reduced to 20 percent. This result is undoubtedly more noisy, however it still provides a general description of the subsurface key structures. The corresponding angle gathers (Figures \ref{fig:anglegather_overthrust_sparsel2}a and \ref{fig:anglegather_overthrust_sparsel2}b) are also of satisfactory quality; this proves once again the effectiveness of the UD-RM method in handling sparse acquisitions within complex geological settings.


\section*{Discussion}
As anticipated in the Introduction section, an interesting connection can be drawn between the UD-RM method and the migration scheme proposed by \cite{Lu2015} to image free-surface multiples in OBS datasets. Given the acquisition geometry of the UD-RM method (Figure \ref{fig:geometries}c), the source wavefield of \cite{Lu2015} can be interpreted as the forward propagation of the recorded down-going data at the source level ($\partial _{z}\mathbf{P}^{+,+}$) with the direct arrival from the source to the focal point (expressed as $\mathbf{G}_\mathbf{d}=\mathbf{f}_\mathbf{d}^{+*}$ in our terminology). Similarly, their receiver wavefield can be represented by the backward propagation of the recorded up-going data at the source level ($\partial _{z}\widetilde{\mathbf{P}}^{+,-}$) with the direct arrival from the source to the focal point. As such their redatumed Green's functions can be expressed as
\begin{equation}
\label{eq:SWIM_s}
\mathbf{g}^{+,+} \approx (\partial _{z}\mathbf{P}^{+,+*} \mathbf{f}_\mathbf{d}^{+*})^{*} = \partial _{z}\mathbf{P}^{+,+}\mathbf{G}_\mathbf{d}
,
\end{equation}
\begin{equation}
\label{eq:SWIM_r}
\mathbf{g}^{+,-} \approx \partial _{z}\widetilde{\mathbf{P}}^{+,-} \mathbf{f}_\mathbf{d}^{+} = \partial _{z}\widetilde{\mathbf{P}}^{+,-}\mathbf{G}_\mathbf{d}^{*}
.
\end{equation}
These wavefields can be also obtained from equation \ref{eq:UDRM_matrix} by assuming an initial knowledge of the focusing functions (i.e., $\mathbf{f}^{+}=\mathbf{f}_\mathbf{d}^{+}$, $\mathbf{f}^{-}=0$); therefore, we can conclude that the wavefields retrieved by \cite{Lu2015} are equivalent to those obtained by our method when ignoring the contributions of free- and internal multiples in the focusing functions. On the other hand, when the UD-RM equations are inverted for the total focusing functions, the retrieved Green's functions correctly account for overburden and free-surface effects, providing a superior input for the subsequent imaging step. This may explain why post-imaging techniques have been developed by \cite{Lu2016} to suppress artefacts in their imaging products due to the imperfect handling of multiple scattering in their wavefield propagation engine. Moreover, it is worth mentioning that the sources used within the imaging algorithm of \cite{Lu2015} are assumed to be located at the free-surface, such that the up- and down-going wavefields degenerate in the same data with opposite polarity due to the reflectivity of free-surface. Since sources are physically placed at depth in real life experiment (i.e., geometry in Figure \ref{fig:geometries}c), the up- and down-going wavefields are effectively different (the down-going component is equal to the up-going component further propagated from the source level to the free-surface and back to the source level). Not taking this into account during redatuming leads to a broader source wavelet in the final redatumed wavefields and corresponding image (because of the auto-correlation of the source wavelet and source ghost), as also acknowledged by \cite{Lu2015}. Being our method based on solid first principles of wave propagation, a recipe for the correct handling of the different components of the total wavefield is naturally revealed in equation equation \ref{eq:UDRM_matrix} .

A current limitations of the proposed method, when compared to, for example, the RM method, is the requirement for wavefield separation to be performed both on the receiver and source sides. Whilst multi-component receivers are always available in modern ocean bottom acquisition systems, dual-sensor sources are still in their infancy (\cite{Halliday2012, Vasconcelos2012, Robertsson2012, Robertsson2008}). As such, source-side deghosting represents nowadays the most likely option to unleash the power of the UD-RM scheme. As acknowledged by \cite{Robertsson2016, Grion2016}, source deghosting remains an unsolved problem in geophysics with few solutions available to date: most of these techniques adopt ideas initially developed for receiver-side deghosting (e.g., \cite{Robertsson2017, Cecconello2019}) and require dense spatial sampling. Whilst a similar requirement holds for the spatial integrals in the UD-RM equations, previous research \citep{Ravasi2020} has shown that the Marchenko methods can successfully operate with mildly aliased data, a situation that is more difficult to handle in the deghosting process. An attractive alternative for the UD-RM method might involve discovering a strategy that eliminates the need for source deghosting. An approach recently developed for source-side multi-dimensional deconvolution \citep{Haacke2023} can circumvent such a pre-processing step; given the similarity of their equation with equation \ref{eq:UD_g}, we envision that a new UD-RM method could be developed with no need for source deghosting.

Moreover, as shown in the Numerical Examples section, the UD-RM method can handle coarsely sampled receiver grids when the associated equations are solved by means of sparsity-promoting inversion; however, the improvements in the retrieved focusing functions, and therefore in the associated Green's functions, may not justify the increased computational cost if the ultimate goal is structural imaging. On the other hand, the uplift in the quality of the recovered angle gathers for both examples seems to suggest the importance of using sparse inversion for subsequent quantitative interpretation analysis. As recently shown in the context of multi-dimensional deconvolution by \cite{Kumar2022} and \cite{Chen2023}, sparsity is not the only viable approach to provide additional prior information to inverse process; a similar observation can be made UD-RM equations when dealing with sparse receiver geometries. Low-rank approximations of the frequency-domain focusing functions could, in fact, represent a more suitable choice that not only regularizes the problem but, at the same time, reduces the dimensionality of the unknown vector. Finally, whilst our current implementation is limited to 2D, recent research in the context of Marchenko redatuming has shown that the extension of these computationally expensive multi-dimensional convolutional methods to 3D is within reach of our current compute capabilities~\citep{Ravasi2021, Brackenhoff2022}. A key component to such a success is represented by our ability to compress the kernels of the convolutional integrals (in our case the down-down-going wavefield) by means of low-rank revealing transforms (e.g., Singular Value Decomposition - SVD) as succesfully demostrated in \cite{ravasimdd21, hong2021,ravasi2022, Yuxi2022IJHPCA, ltaif2023}.

\section*{Conclusions}
We have presented a novel Marchenko-type redatuming scheme, named the upside-down Rayleigh-Marchenko method. Being based on a similar mathematical derivation to that of the RM method, the UD-RM method shares all the benefits of its predecessor; these include relaxed requirements in terms of source wavelet estimation and pre-processing and positioning of sources and receivers, as well as the ability to handle both free-surface and internal multiples. Moreover, by invoking reciprocity, the UD-RM method can handle sparse receiver geometries as all spatial integrals are performed over source arrays, making it a suitable candidate for accurate redatuming of OBS acquisition systems. Similar to other mirror migration algorithm, UD-RM only considers the down-going component of the recorded wavefield at the receiver array: this implies that receivers are mirrored with respect to the sea surface, potentially ‘expanding’ the illumination of the retrieved Green’s functions compared to those obtained by other Marchenko redatuming schemes, such as the RM method. However, in order to obtain the input data for the proposed method, two consecutive data processing steps are required. In the presence of dual-source, multi-component data, the PZ summation method can be used to separate the data both on the source and the receiver sides. Alternatively, a step of model-based source deghosting must be performed in the presence of single-source, multi-component data. Given the added algorithmic complexity of source deghosting compared to data-driven wavefield separation methods, relaxing the acquisition requirements from dual-source to single-source data is shown to lead to the introduction of small artifacts in the reconstructed wavefields. It is however worth mentioning that, whilst the equations derived in this paper indicate that the input data is the separated wavefield from dipole sources, we have validated that utilizing the separated wavefield from monopole sources as input yields identical results.
Finally, whilst the UD-RM equations have been shown to be well posed and invertible by means of least-squares solvers in ideal acquisition scenarios, numerical results in the presence of sparse receiver arrays reveal the need for more advanced solvers. To limit the rank-deficient nature of the problem when the number of sources far exceeds that of receivers, sparsity-promoting inversion coupled with a sliding linear Radon transform is shown to provide more accurate focusing functions that those from least-squares inversion. 

\section{Acknowledgments}
The authors thank KAUST for supporting this research. All numerical examples have been developed using the PyLops ramework \citep{Ravasi2020} and the associated PyMarchenko library (https://github.com/DIG-Kaust/pymarchenko).

\newpage
\clearpage

\appendix

\section{Appendix A}
Since $g_{0}(\textbf{x}_{R},\textbf{x}_{S}')$ on the left-hand side of equation \ref{eq:g_g0} contains all those events that reach the receiver level without having experienced any interaction with the free-surface, it can be expressed as the sum of the direct wave (i.e, the only down-going component of the wavefield in a medium without free-surface) and the up-going wavefield at both $\textbf{x}_{R}$ and $\textbf{x}_{S}'$:
\begin{equation}
\label{eq:g0_decomposation}
g_{0}(\textbf{x}_{R},\textbf{x}_{S}') = g_{d}(\textbf{x}_{R},\textbf{x}_{S}') + g_{0}^{-,-}(\textbf{x}_{R},\textbf{x}_{S}'),
\end{equation}
where the second term on the right-hand side of equation \ref{eq:g0_decomposation} can be expanded as:
\begin{equation}
\label{eq:g--_decomposation}
g_{0}^{-,-}(\textbf{x}_{R},\textbf{x}_{S}') = g^{-,-}(\textbf{x}_{R},\textbf{x}_{S}') - g_{fs}^{-,-}(\textbf{x}_{R},\textbf{x}_{S}'),
\end{equation}
where $g^{-,-}(\textbf{x}_{R},\textbf{x}_{S}')$ is the total up-going wavefield at both $\textbf{x}_{R}$ and $\textbf{x}_{S}'$ from medium with free-surface, while $g_{fs}^{-,-}(\textbf{x}_{R},\textbf{x}_{S}')$ refers to the part of such a wavefield that has experienced at least one interaction with the free-surface. Combining equations \ref{eq:g0_decomposation} and \ref{eq:g--_decomposation}, the left-hand side of equation \ref{eq:g_g0} can be expressed as:
\begin{equation}
\begin{aligned}
\label{eq:g_g0_lhs}
&g^{.,-}(\textbf{x}_{R},\textbf{x}_{S}') - g_{0}(\textbf{x}_{R},\textbf{x}_{S}')
\\=&g^{.,-}(\textbf{x}_{R},\textbf{x}_{S}') - (g_{d}(\textbf{x}_{R},\textbf{x}_{S}') + g^{-,-}(\textbf{x}_{R},\textbf{x}_{S}') - g_{fs}^{-,-}(\textbf{x}_{R},\textbf{x}_{S}'))
\\=&g^{.,-}(\textbf{x}_{R},\textbf{x}_{S}') - g^{-,-}(\textbf{x}_{R},\textbf{x}_{S}') - g_{d}(\textbf{x}_{R},\textbf{x}_{S}') + g_{fs}^{-,-}(\textbf{x}_{R},\textbf{x}_{S}')
\\=&g^{+,-}(\textbf{x}_{R},\textbf{x}_{S}') - g_{d}(\textbf{x}_{R},\textbf{x}_{S}') + g_{fs}^{-,-}(\textbf{x}_{R},\textbf{x}_{S}')
\end{aligned}
,
\end{equation}
where the term $g_{fs}^{-,-}(\textbf{x}_{R},\textbf{x}_{S}')$ can be expressed using the following convolutional integral relation:
\begin{equation}
\label{eq:g_fs}
g_{fs}^{-,-}(\textbf{x}_{R},\textbf{x}_{S}') = -\int _{\Lambda _{S}}\dfrac{2}{j\omega \rho(\textbf{x}_{S})}\partial _{z}g^{-,+}(\textbf{x}_{R},\textbf{x}_{S})g_{0}(\textbf{x}_{S},\textbf{x}_{S}')d\textbf{x}_{S}.
\end{equation}
Substituting equation \ref{eq:g_fs} into the left-hand side of equation \ref{eq:g_g0}, and moving $g_{fs}^{-,-}(\textbf{x}_{R},\textbf{x}_{S}')$ on the other side of the equal leads to:
\begin{equation}
\begin{aligned}
\label{eq:UD_g_derivation}
&g^{+,-}(\textbf{x}_{R},\textbf{x}_{S}') - g_{d}(\textbf{x}_{R},\textbf{x}_{S}') 
\\=&-\int _{\Lambda _{S}}\dfrac{2}{j\omega \rho(\textbf{x}_{S})}\partial _{z}g^{.,+}(\textbf{x}_{R},\textbf{x}_{S})g_{0}(\textbf{x}_{S},\textbf{x}_{S}')d\textbf{x}_{S} 
\\&-g_{fs}^{-,-}(\textbf{x}_{R},\textbf{x}_{S}')
\\=&-\int _{\Lambda _{S}}\dfrac{2}{j\omega \rho(\textbf{x}_{S})}\partial _{z}g^{.,+}(\textbf{x}_{R},\textbf{x}_{S})g_{0}(\textbf{x}_{S},\textbf{x}_{S}')d\textbf{x}_{S} 
\\&+\int _{\Lambda _{S}}\dfrac{2}{j\omega \rho(\textbf{x}_{S})}\partial _{z}g^{-,+}(\textbf{x}_{R},\textbf{x}_{S})g_{0}(\textbf{x}_{S},\textbf{x}_{S}')d\textbf{x}_{S}
\\=&-\int _{\Lambda _{S}}\dfrac{2}{j\omega \rho(\textbf{x}_{S})}\partial _{z}g^{+,+}(\textbf{x}_{R},\textbf{x}_{S})g_{0}(\textbf{x}_{S},\textbf{x}_{S}')d\textbf{x}_{S}
\end{aligned}
\end{equation}
which is equivalent to equation \ref{eq:UD_g}.

\bibliographystyle{seg}  
\bibliography{example}

\end{document}